\documentclass[twocolumn,showpacs,preprintnumbers,nofootinbib,prd,
superscriptaddress,10pt]{revtex4-2}

\pdfoutput=1

\usepackage{amsmath,amssymb}
\usepackage{amsfonts}
\usepackage{mathtools}
\usepackage[normalem]{ulem}
\usepackage{textcomp}
\usepackage{enumitem}
\usepackage{bm}
\usepackage{bbm}
\usepackage{afterpage}
\usepackage{graphicx}
\graphicspath{{img}} 
\usepackage{tensor}

\usepackage{tabularx}
\usepackage{longtable}
\usepackage{makecell}
\usepackage{multirow}
\usepackage{booktabs}

\usepackage{layouts}
\usepackage[usenames,dvipsnames]{xcolor}
\usepackage[utf8]{inputenc}
\usepackage{rotating}
\usepackage[bookmarks, pdfusetitle, pdfauthor={Stefano Schmidt}]{hyperref}
\usepackage{blindtext}
\usepackage{graphicx}
\usepackage{siunitx}
\usepackage{orcidlink}

\renewcommand{\d}[1]{\ensuremath{\operatorname{d}\!{#1}}}

\newcommand{\scalar}[2]{\langle #1|#2 \rangle}

\newcommand{\rescalar}[2]{( #1 |#2 )}

\newcommand{\imscalar}[2]{[ #1|#2 ]}

\begin{document}

\begin{abstract}
	We propose a novel signal-consistency test applicable to a broad search for gravitational waves emitted by generic binary black hole (BBH) systems. The test generalizes the time domain $\xi^2$ signal-consistency test currently utilized by the GstLAL pipeline, which quantifies the discrepancy between the expected signal-to-noise ratio timeseries with the measured one. While the traditional test is restricted to aligned-spin circular orbits and does not account for higher-order modes (HMs), our test does not make any assumption on the nature of the signal.
	After addressing the mathematical details of the new test, we quantify its advantages in the context of searching for precessing BBHs and/or BBHs with HM content. Our results reveal that for precessing signals, the new test is optimal and has the potential to reduce the values of the $\xi^2$ statistics by up to two orders of magnitude when compared to the standard test. However, in the case of signals with HM content, only a modest enhancement is observed.
	Recognizing the computational burden associated with the new test, we also derive an approximated signal-consistency test. This approximation maintains the same computational cost as the standard test and can be easily implemented in any matched filtering pipeline with minimal changes, sacrificing only a few percent of accuracy in the low SNR regime. However in the high SNR regime the approximated signal consistency test does not bring any improvement as compared to the ``standard'' one.
	By introducing our new test and its approximation and understanding their validity and limitation, this work will benefit any matched-filtering pipeline aimed at searching for BBH signals with strong precession and/or HM content.
\end{abstract}
	
 \title{A novel signal-consistency test for gravitational-wave searches of generic black hole binaries}
	\author{Stefano \surname{Schmidt} \orcidlink{0000-0002-8206-8089}}
		\email{s.schmidt@uu.nl}
        \affiliation{Nikhef, Science Park 105, 1098 XG, Amsterdam, The Netherlands}
        \affiliation{Institute for Gravitational and Subatomic Physics (GRASP),
Utrecht University, Princetonplein 1, 3584 CC Utrecht, The Netherlands}
	\author{Sarah \surname{Caudill} \orcidlink{0000-0002-8927-6673}}
		\affiliation{Department of Physics, University of Massachusetts, Dartmouth, MA 02747, USA}
		\affiliation{Center for Scientific Computing and Data Science Research, University of Massachusetts, Dartmouth, MA 02747, USA}
	\maketitle

\section{Introduction}

Gravitational-wave searches for binary black holes (BBHs) have been a cornerstone in the activities of the LIGO-Virgo-KAGRA (LVK) collaboration~\cite{LIGOScientific:2014pky, VIRGO:2014yos, KAGRA} and have made possible the discovery of nearly one hundred GW events during the first three observing runs \cite{LIGOScientific:2018mvr, LIGOScientific:2020ibl, LIGOScientific:2021usb, KAGRA:2021vkt}.

BBH searches can either be performed in a model-dependent way through the technique of 
{\it matched filtering} \cite{Sathyaprakash:1991mt, Dhurandhar:1992mw, Allen:2005fk, Cannon:2011vi, Babak:2012zx} or a model-independent way through excess-power methods \cite{Klimenko:2008fu, Necula:2012zz, Drago:2020kic}. The matched-filter method computes the correlation between GW detector data and a set of templates of modelled BBH waveforms. This method proves more sensitive for lower-mass systems, as the power in these signals is spread over many time-frequency bins \cite{LIGOScientific:2016sjg, LIGOScientific:2017vwq}. Beyond the initial matched-filter or excess-power stage, searches consist of analysis pipelines \cite{Privitera:2013xza, Adams:2015ulm, Usman:2015kfa, Capano:2016dsf, Messick:2016aqy, Nitz:2017svb, gstlal_paper2, Hanna:2019ezx, Aubin:2020goo, Davies:2020tsx, Chu:2020pjv, Ewing:2023qqe} that employ various statistical tests to improve the separation of GW candidates from false alarms caused by the detectors' non-stationary and non-Gaussian noise \cite{Blackburn:2008ah, LIGOScientific:2016gtq, Cabero:2019orq, LIGO:2020zwl, LIGO:2021ppb}.

A class of these statistical tests, known as signal-consistency tests \cite{Allen:2004gu, Shawhan:2004qq, Cannon:2015gha, Messick:2016aqy, Nitz:2017lco, Dhurandhar:2017aan, Gayathri:2019omo, Hanna:2019ezx, Godwin:2020weu, McIsaac:2022odb, Tsukada:2023edh}, have been designed to further distinguish between false alarms and real GW signals by computing the agreement between the data and the signal model assumed by the search.
Signal-consistency tests have been implemented by all LVK matched-filtering pipelines and have played an integral role in enabling many high-significance GW detections.
%
%

Expanding our searches to include signals from precessing binaries \cite{DalCanton:2014qjd, PhysRevD.89.024010, PhysRevD.102.041302, Indik:2016qky, Harry:2016ijz, McIsaac:2023ijd} and/or binaries with higher-order modes (HMs) \cite{CalderonBustillo:2015lrt, Harry:2017weg, Chandra:2022ixv, 2021PhRvD.103b4042M, Wadekar:2023kym} necessitates the generalization of signal-consistency tests to a more versatile framework. Neglecting to update these tests can result in decreased search sensitivity, potentially offsetting the benefits of using a more diverse set of templates.

While the $\chi^2$ time-frequency signal-consistency test \cite{Allen:2004gu} and its variant, the sine-Gaussian $\chi^2$ discriminator~\cite{Nitz:2017lco}, have been successfully applied in searches including higher-order modes \cite{Harry:2017weg, Chandra:2022ixv} and precessing signals \cite{McIsaac:2023ijd}, little attention has been given to generalizing other types of signal-consistency tests, or even to the development of new ones.

In particular, in this work we focus on the autocorrelation-based least-squares test, denoted $\xi^2$, which is currently utilized by the GstLAL search pipeline \cite{Messick:2016aqy, gstlal_paper2, cannon2020gstlal, Ewing:2023qqe}.
The test relies on the assumption that the signal to detect is a circular aligned-spin binary system where no HMs are considered. In this scenario, the system is symmetric for rotation around the orbital rotation axis and this symmetry traslates into a simple relation between the two polarizations of the GW emitted, which allows to obtain a convenient and computationally efficient expression for the test.

In this work, we drop the ``aligned-spin and no HM'' assumption and we introduce a new signal-consistency test that does not make {\it any assumption} about the nature of the signal to detect. Thus, while the test was primarly motivated to search for precessing and/or HM signals, it can be applied to a matched filtering search for any type of gravitational waves signal.
In Sec.~\ref{sec:background}, we provide some general background on matched filtering and on the state of the art signal-consistency test, while in Sec.~\ref{sec:chisq_sym} we introduce our new generalized signal-consistency test $\xi_\text{sym}^2$ and provide a computationally convenient approximate expression $\xi_\text{mix}^2$ for it.
In Sec.~\ref{sec:validation} we discuss the performance of the newly introduced test and its approximated version.
Sec.~\ref{sec:conclusion} gathers some final remarks.

\section{Background}\label{sec:background}

According to the theory of general relativity \cite{Maggiore:2007ulw}, a gravitational wave only carries two physical degrees of freedom $h_+$ and $h_\times$, also called {\it polarizations}.
For a generic GW emitted by a compact binary, the polarizations depend on the {\it intrinsic} properties of the binary such as the two compact objects' masses $m_1$, $m_2$ and spins $\mathbf{s}_1$, $\mathbf{s}_2$. The signal observed at the source also depends on the so-called {\it extrinsic} properties including position, usually parametrized in spherical coordinates by a distance $D$ from the origin, a polar angle called {\it inclination} angle $\iota$, and an azimuthal angle $\phi$.

For the purpose of modeling, it is customary to expand the gravitational waveform's dependence on the extrinsic parameters in terms spin-2 spherical harmonics $Y_{\ell m}(\iota, \phi)$ \cite{Maggiore:2007ulw}\:
\begin{equation}
	h_+ + ih_\times = \frac{1}{D} \sum^{\infty}_{\ell = 0} \sum^{\ell}_{m = -\ell} Y_{\ell m}(\iota, \phi) e^{i m \phi}h_{\ell m}(t)
\end{equation}
where the {\it modes} $h_{\ell m}$ are complex functions of the intrinsic parameters $m_1, m_2, \mathbf{s}_1, \mathbf{s}_2$.
Each mode can be decomposed into a time-dependent amplitude $A_{\ell m}$ and a time-dependent phase $\varphi_{\ell m}$ such that:
\begin{equation}\label{eq:amp_phase_decomposition}
	h_{\ell m} = A_{\ell m} e^{i\varphi_{\ell m}}.
\end{equation}
As a consequence the imaginary part $h^\text{I}_{\ell m}$ of each mode is equivalent to the real part $h^\text{R}_{\ell m}$ shifted by a constant phase of $\pi/2$. In the frequency domain, this takes the simple form of
\begin{equation}\label{eq:real_imag_symmetry}
	\tilde{h}^\text{R}_{\ell m} = i \tilde{h}^\text{I}_{\ell m}
\end{equation}
where $\tilde{\phantom{a}}$ indicates the Fourier transform.

If the two spins are aligned with $\mathbf{L}$, the orbital plane will point toward a fixed direction, establishing an axis of symmetry for the non-precessing binary system. Mathematically, this symmetry translates into a symmetry between the modes:
\begin{equation}\label{eq:pm_mm_symmetry}
	h_{\ell m} = (-1)^{\ell} h^*_{\ell -m}
\end{equation}
where $*$ denotes complex conjugation.
On the contrary, if the two spins $\mathbf{s}_1, \mathbf{s}_2$ are misaligned with the binary orbital angular momentum $\mathbf{L}$, the binary plane experiences precessional motion, where the orbital angular momentum rotates around a (roughly) constant direction \cite{Apostolatos:1994mx, Kidder:1992fr, Kidder:1995zr, Buonanno:2002fy, Campanelli:2006fy} and the symmetry Eq.~\eqref{eq:pm_mm_symmetry} is no longer valid.

In most aligned-spins systems, it turns out that only the $\ell = |m| = 2$ modes give a significant contribution to the polarizations and thus all the other HMs are neglected \cite{Maggiore:2007ulw, Pekowsky:2012sr, Healy:2013jza, CalderonBustillo:2015lrt}.
Thus, the waveform from a non-precessing binary without imprints from HMs has a strikingly simple expression:
\begin{align}
	 h_+ &= \frac{1}{D} \frac{1+\cos^2 \iota}{2} \; \Re \lbrace h_{22}e^{i2\phi}\rbrace \label{eq:wf_22_p},\\
	 h_\times &= \frac{1}{D} \cos \iota \; \Im \lbrace h_{22}e^{i2\phi} \rbrace \label{eq:wf_22_c}.
\end{align}
Note that as a consequence of Eq.~\eqref{eq:real_imag_symmetry}, the two polarizations in Fourier space are related by the simple relation
\begin{equation}\label{eq:symmetry_polarization_plus_cros}
	\tilde{h}_+ \propto i \tilde{h}_\times
\end{equation}
and, for $\iota = 0$, we have trivially $\tilde{h}_+ = i \tilde{h}_\times$.

In the remainder of this section, we describe how the simplicity of Eqs.~\eqref{eq:wf_22_p} and \eqref{eq:wf_22_c} enters the the expression for the currently used $\xi^2$ signal-consistency test \cite{Messick:2016aqy} and we describe how to move away from the assumption of aligned-spin systems without HMs, tackling the most general case.

\subsection{Overview}

The core of matched filter relies on computing the cross-correlation between two time-series $a(t), b(t)$, weighted by the Power Spectral Density (PSD) of the noise $S_n(f)$. Mathematically, we can compute this cross-correlation by defining a time-dependent {\it complex} scalar product:
\begin{equation}\label{eq:time_dependent_scalar_product}
	\scalar{a}{b}(t) = 4 \int_{0}^{\infty} \d{f} \; \frac{\tilde{a}^{*}(f) \tilde{b}(f)}{S_n(f)} e^{-i2\pi ft}
\end{equation}
It is convenient to separate the real $\rescalar{\cdot}{\cdot}$ and imaginary $\imscalar{\cdot}{\cdot}$ part of the scalar product as:
\begin{equation}
	\scalar{a}{b}(t) = \rescalar{a}{b}(t) + i \imscalar{a}{b}(t)
\end{equation}
Given a timeseries $a(t)$, we can use the above scalar product to computed the normalized timeseries $\hat{a}(t)$:
\begin{equation}
	\hat{a}(t) = \frac{a(t)}{\sqrt{\rescalar{a}{a}(t=0)}}
\end{equation}

Following the notation of \cite{Messick:2016aqy}, the output of the matched-filtering procedure is a complex timeseries $z(t)$:
\begin{equation}
	z(t) = \rescalar{d}{h_R}(t) + i\rescalar{d}{h_I}(t)
\end{equation}
where $h_R$ and $h_I$ are two {\it normalized} real templates.
We may also define the Signal-to-noise ratio (SNR) timeseries $\rho(t)$ as:
\begin{equation}\label{eq:snr_def}
	\rho(t) = |z(t)| = \sqrt{\rescalar{d}{h_R}^2(t) + \rescalar{d}{h_I}^2(t)}
\end{equation}
The {\it real} filters $h_R, h_I$ are chosen to maximise $\rho$ at the time where a GW signal is present in the data and their expression depends on the assumptions about the nature of the GW signal to search.
For instance, if a template is spin-aligned and HM are not considered, it is sufficient to filter the data with the two polarizations $\hat{h}_+, \hat{h}_\times$ evaluated at zero inclination $\iota = 0$. As will be shown below, a more complicated expression will be needed for a more general case.

Given a trigger at time $t=0$, the $\xi^2$ test relies on predicting the SNR timeseries $z(t)$ obtained by filtering a signal $h$ with a matching templates. The predicted timeseries $R(t)$ is then compared to the measured timeseries $z(t)$ to compute the squared residual timeseries:
\begin{equation}\label{eq:chisq_timeseries}
	\xi^2(t) = |z(t)-R(t)|^2
\end{equation}
We can integrate the residual timeseries to obtain the $\xi^2$ statistics:
\begin{equation}\label{eq:chisq}
	\xi^2 = \frac{\int_{-\delta t}^{\delta t} \; \d{t} \;\; |z(t)-R(t)|^2}{\int_{-\delta t}^{\delta t} \; \d{t} \;<\xi^2(t)>}
\end{equation}
where the integral extends on a short time window $[-\delta t, \delta t]$ around the trigger time.
To obtain the $\xi^2$ statistics integral of the residual timeseries is normalized by integral of the expected value $<\xi^2(t)>$ over different Gaussian noise realizations of the residual timeseries without a signal buried in the noise. Clearly, its value depends on the the templates employed.
It is convenient to express $\delta t$ in terms of the so-called autocorrelation length (ACL) \cite{Messick:2016aqy}, defined as the number of samples in the time-window $[-\delta t, \delta t]$ at a given sample rate $f_\text{sampling}$, so that $\delta t = (\textrm{ACL}-1)/2\, f_\text{sampling}$.


The $\xi^2$ defined above can be used by the GW search pipelines to veto some loud triggers. If a trigger is caused by a noise fluctuation or non-Gaussian noise transient bursts \cite{Blackburn:2008ah}, the discrepancy between the expected and measured SNR timeseries will be large, leading to a large value of $\xi^2$. This can be used to downrank certain triggers, with large improvement in sensitivity.

As it is custom, to predict the SNR timeseries we model the data $d$ as a superposition of Gaussian noise $n$ and a GW signal $h$: $d = n + h$.
For current ground based interferometers, the GW signal is:
\begin{equation}\label{eq:strain}
	h = F_+ h_+ + F_\times h_\times
\end{equation}
where $F_+, F_\times$ are the Antenna Pattern Functions \cite{Finn:1992xs, Jaranowski:1998qm}, which define the detector's response to a signal coming from a given direction in the sky.
For our purpose, it is convenient to express the signal model in terms of the {\it normalized} polarizations:
\begin{equation}\label{eq:strain_normalized}
	h = \mathcal{F}_+ \hat{h}_+ + \mathcal{F}_\times \hat{h}_\times
\end{equation}
where we absorbed into $\mathcal{F}_+, \mathcal{F}_\times$ an overall scaling factor (which depends on the source distance and on the sky location).
Of course, $\mathcal{F}_+, \mathcal{F}_\times$ are not known at the moment of the search but must be inferred from the value of the SNR timeseries $z(0)$ at the time of a trigger.

\subsection{The original $\xi^2$ test} \label{sec:chisq_std}

To obtain an expression for the filters $h_R, h_I$ to use in the case of a spin-aligned system where HM are not considered, we must first write the signal model Eq.~\eqref{eq:strain_normalized} in a convenient way~\cite{Allen:2005fk, Harry:2016ijz}:
\begin{equation}\label{eq:NP_signal}
	h = \mathcal{A} \; \Re \lbrace h_{22} e^{i\phi_C} \rbrace = \mathcal{A} \, \left( \hat{h}^\text{R}_{22} \cos\phi_C - \hat{h}^\text{I}_{22} \sin\phi_C \right)
\end{equation}
where $\mathcal{A}$ is an amplitude factor and $\phi_C$ is a constant phase shift, both depending on $\mathcal{F}_+, \mathcal{F}_\times, \iota$ and $\phi$ and we used Eqs.~\eqref{eq:wf_22_p} and \eqref{eq:wf_22_c} for the polarizations.
Using Eq.~\eqref{eq:real_imag_symmetry}, the frequency domain signal model takes a remarkably simple form of:
\begin{equation}\label{eq:NP_signal_FD}
	\tilde{h} = \mathcal{A} e^{i\phi_C} \; \tilde{\hat{h}}^\text{R}_{22}
\end{equation}
In practise this means that all the variability of the observed signals is encoded in the real function $\hat{h}^\text{R}_{22}(t)$ and that all the possible effects due to inclination, reference phase and sky localization only affect an overall amplitude and phase.

With equation Eq.~\eqref{eq:NP_signal_FD} at hand, it is a simple exercise to show that the SNR timeseries is maximised by the following filters
\begin{equation}\label{eq:templates_std}
\begin{aligned}
	&h_R = \hat{h}^\text{R}_{22} \\
	&h_I = \hat{h}^\text{I}_{22}
\end{aligned}
\end{equation}
and thus that the search statistics $\rho(t)$ does not depend on $\iota$ and $\phi$.
We call ``standard'' the SNR timeseries obtained using such templates:
\begin{equation}\label{eq:std_snr}
	\rho_\text{std}(t) = \sqrt{\rescalar{h}{\hat{h}^\text{R}_{22}}^2 + \rescalar{h}{\hat{h}^\text{I}_{22}}^2}
\end{equation}
Note that in this simple case, the two templates in frequency domain exibith a remarkable symmetry, inherited from Eq.~\eqref{eq:real_imag_symmetry}:
\begin{equation}\label{eq:symmetry_22_real_imag_templates}
	\tilde{h}_R = i \tilde{h}_I
\end{equation}
While the ``standard'' signal consistency test relies on this important symmetry, in general Eq.~\eqref{eq:symmetry_22_real_imag_templates} is not true for a precessing and/or HM signal: this is the main motivation for the novel signal consistency test introduced in this work.

Thanks to this remarkable property of the signal, the predicted SNR timeseries $R_\text{std}(t)$ around a trigger at $t = 0$ is given by:
\begin{align}
	R_\text{std}(t) &= \rescalar{h}{\hat{h}^\text{R}_{22}}(t) + i \rescalar{h}{\hat{h}^\text{I}_{22}}(t) \nonumber \\
	&= z(0) \, \Big\lbrace\rescalar{\hat{h}^\text{R}_{22}}{\hat{h}^\text{R}_{22}}(t) + i \rescalar{\hat{h}^\text{R}_{22}}{\hat{h}^\text{I}_{22}}(t)\Big\rbrace
	 \label{eq:R_std}
\end{align}
where, to simplify the expression above, we used the fact that $h = \mathcal{A} e^{i\phi_C} \hat{h}^\text{R}_{22}$ and that $z(0) = \mathcal{A} \, e^{-i\phi_C}$, since $\rescalar{\hat{h}^\text{I}_{22}}{\hat{h}^\text{R}_{22}}(t=0)$ is equal to zero.
By comparing the measured timeseries $z(t)$ and the expected timeseries $R_\text{std}(t)$ with Eq.~\eqref{eq:chisq}, one can compute $\xi_\text{std}^2$ for a non-precessing search.
It is a simple exercise to compute the expected value of the residual timeseries over different noise realizations~\cite[App.~A]{Messick:2016aqy}:
\begin{align}
	<\xi_\text{std}^2&(t)> = 2 - 2\left|\rescalar{\hat{h}^\text{R}_{22}}{\hat{h}^\text{R}_{22}}(t) + i\rescalar{\hat{h}^\text{R}_{22}}{\hat{h}^\text{I}_{22}}(t)\right|^2  \label{eq:expected_value_std}
\end{align}
The integral of this expression can be used as a normalization for the $\xi^2$ statistics.

By using again Eq.~\eqref{eq:symmetry_22_real_imag_templates} and the identity $\imscalar{a}{ib} = \rescalar{a}{b}$, the ``standard'' SNR Eq.~\eqref{eq:std_snr} can also be expressed in terms of $\hat{h}^\text{R}_{22}$ only:
\begin{equation}\label{eq:std_snr_bis}
	\rho_\text{std}(t) = \sqrt{\rescalar{h}{\hat{h}^\text{R}_{22}}^2 + \imscalar{h}{\hat{h}^\text{R}_{22}}^2}
\end{equation}
Of course the same is readily done for the $R_\text{std}(t)$:
\begin{align}
	R_\text{std}(t) &= z(0) \Big\lbrace \rescalar{\hat{h}^\text{R}_{22}}{\hat{h}^\text{R}_{22}}(t) + i \imscalar{\hat{h}^\text{R}_{22}}{\hat{h}^\text{R}_{22}}(t) \Big\rbrace \nonumber \\
		&= z(0) \, \scalar{\hat{h}^\text{R}_{22}}{\hat{h}^\text{R}_{22}}(t)\label{eq:R_std_bis}
\end{align}
The quantity $\scalar{\hat{h}^\text{R}_{22}}{\hat{h}^\text{R}_{22}}(t)$ is sometimes called template autocorrelation.
For ``standard'' signals, the Eqs.~\eqref{eq:R_std} and~\eqref{eq:R_std_bis} are equivalent.
However, for precessing and/or HM they may give very different results, due to the breaking of the symmetry Eq.~\eqref{eq:symmetry_22_real_imag_templates}.

We close by noting that the predicted timeseries is given by a product of a trigger-dependent scalar and a complex template-dependent timeseries
\begin{equation*}\label{eq:simple_factorization}
	R(t) = \text{Trigger} \times \text{Complex timeseries}
\end{equation*}
This arises directly from the fact that the signal model Eq.~\eqref{eq:NP_signal_FD} presents the same convenient factorization.
Such factorization makes the $\xi^2$ evaluation particularly computationally convenient and hence the $\xi^2$ test computationally attractive.
As we will see, the use of precessing and/or HM templates breaks this factorization, as Eq.~\eqref{eq:NP_signal_FD} is no longer valid.


%
%
%
%
%
%

\begin{figure*}[t!]
	\includegraphics{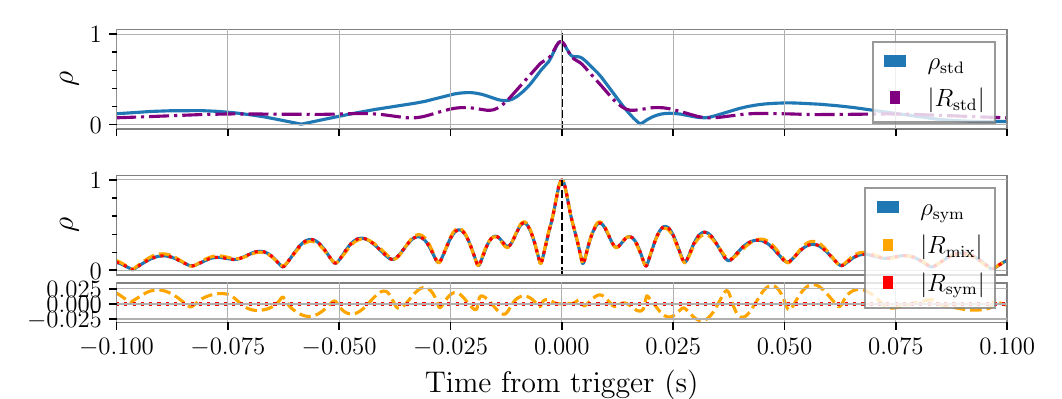}
	\caption{
	Predicted and measured absolute value of the SNR timeseries, $|R|$ and $\rho$ respectively, for a precessing signal in zero noise and with unit magnitude, filtered with a perfectly matching template.
	We measure the SNR using both the ``standard'' SNR and the ``symphony" SNR. The predicted SNR is computed with three different prescriptions, $R_\text{std}$ Eq.~\eqref{eq:R_std_bis} suitable for the ``standard'' SNR, and $R_\text{sym}$ Eq.~\eqref{eq:R_sym_amazing} and $R_\text{mix}$ Eq.~\eqref{eq:R_mix}, both suitable for the ``symphony" SNR. We also report in the bottom panel the difference between the expected and measured SNR timeseries.
	In the bottom panel we report the difference between both $|R_\text{sym}|$ or $|R_\text{mix}|$ and the ``symphony" SNR. It is manifest that $|R_\text{sym}|$ and $\rho_\text{sym}$ show perfect agreement between each other while $|R_\text{std}|$ and $|R_\text{mix}|$ do not accurately predict the relevant SNR timeseries. Also note that the peak of ``standard'' search statistics is lower than unity, meaning that performing matched filtering with templates Eq.~\eqref{eq:templates_std} is not able to fully recover the SNR of a precessing signal.
	The signal is injected into Gaussian noise, sampled from the PSD LIGO Livingston PSD \cite{O3_PSDs} with a rate of $\SI{4096}{Hz}$. The waveform is characterized by masses $m_1, m_2 = \SI{28}{M_\odot}, \SI{3}{M_\odot}$ and spins $\mathbf{s}_1 = (-0.8, 0.02, -0.5)$ and $\mathbf{s}_2 = 0$, observed with an inclination $\iota = 2.66$. It was generated starting from a frequency of $\SI{12}{Hz}$ with the approximant \texttt{IMRPhenomXP} \cite{Pratten:2020ceb}.
	}
	\label{fig:snr_timeseries}
\end{figure*}

\section{A New Generalized $\xi^2$ Signal-consistency Test} \label{sec:chisq_sym}

The simplicity of the ``standard'' case arises directly from the symmetry between the two polarizations Eq.~\eqref{eq:symmetry_polarization_plus_cros}, which allows us to conveniently factorize the signal model and the predicted SNR timeseries.
However, in general this is not possible and a different expression for the filters and the expected SNR timeseries needs to be computed.
With the {\it only} assumptions that the observed signal is a linear combination of the two polarizations, the appropriate filters for the interferometric data are given by~\cite{Capano:2013raa, Schmidt:2014iyl, Harry:2017weg}:
\begin{equation}\label{eq:templates_sym}
\begin{aligned}
	&h_R = \hat{h}_+ \\
	&h_I = \hat{h}_\perp = \frac{1}{\sqrt{1-\hat{h}_{+\times}^2}} (\hat{h}_\times - \hat{h}_{+\times} \hat{h}_+)
\end{aligned}
\end{equation}
where 
\begin{equation}\label{eq:h_pc_def}
	\hat{h}_{+\times} = \rescalar{\hat{h}_+}{\hat{h}_\times}(t=0) 
\end{equation}
The real quantity $\hat{h}_{+\times}$ is a crucial measure of the precession and/or HM content of a template. The non-precessing non-HM limit can be recovered by $\hat{h}_{+\times}=0$.
We call {\it symphony SNR}\footnote{
The term ``symphony" comes from the title of a paper describing the statistics \cite{Harry:2017weg}.}
the timeseries Eq.~\eqref{eq:snr_def} produced with the templates above:
\begin{equation}\label{eq:sym_snr}
	\rho_\text{sym}(t) = \sqrt{\rescalar{h}{\hat{h}_+}^2 + \rescalar{h}{\hat{h}_\perp}^2}
\end{equation}

Note that if $h_+$ and $h_\times$ are normalized, the template $\hat{h}_\perp$ is normalized by definition: $\scalar{\hat{h}_\perp}{\hat{h}_\perp} = 1$.
Moreover, $\hat{h}_+$ and $\hat{h}_\perp$ are orthogonal vectors, i.e. $\rescalar{\hat{h}_+}{\hat{h}_\perp} = 0$.
Indeed, the vectors for $\hat{h}_+, \hat{h}_\perp$ follows the Gram-Schmidt ``orthogonalization" prescription to create a set of orthonormal basis from the set of basis vectors $\{\hat{h}_+, \hat{h}_\times\}$. Obviously, the fact that the two filters are orthogonal doesn't mean that in general $\hat{h}_+$ and $\hat{h}_\perp$ are related by a simple expression such as Eq.~\eqref{eq:symmetry_22_real_imag_templates}.

To compute the predicted timeseries $R_\text{sym}(t)$, it is convenient to rewrite the signal model Eq.~\eqref{eq:strain_normalized} in terms of $\hat{h}_+$ and $\hat{h}_\perp$
\begin{equation}\label{eq:strain_normalized_symphony}
	h = \mathcal{A}_+ \hat{h}_+ + \mathcal{A}_\perp \hat{h}_\perp
\end{equation}
where $\mathcal{A}_+ = \mathcal{F}_+ + \hat{h}_{+\times} \mathcal{F}_\times$ and $\mathcal{A}_\perp =  \mathcal{F}_\times \sqrt{1 - \hat{h}^2_{+\times}}$.
With this definition, the predicted SNR timeseries is given by:
\begin{align}
	R_\text{sym}(t) =& \rescalar{h}{\hat{h}_+}(t) + i \rescalar{h}{\hat{h}_\perp}(t) \nonumber \\
		=&\; \mathcal{A}_+ \hat{h}_{++}(t) + \mathcal{A}_\perp \hat{h}_{\perp+}(t) \nonumber \\
			&+ i \mathcal{A}_+ \hat{h}_{+\perp}(t) + i \mathcal{A}_\perp \hat{h}_{\perp\perp}(t) \label{eq:R_sym}
\end{align}
where to shorten the notation we defined
\begin{equation}
	\hat{h}_{\bullet\star}(t) = \rescalar{\hat{h}_\bullet}{\hat{h}_\star}(t) \;\;\; \text{with} \;\; \bullet\star = +, \times, \perp
\end{equation}
and we identify $\mathcal{A}_{+/\perp}$ with the real and imaginary part respectively of the trigger $z(0)=\mathcal{A}_+ + i \mathcal{A}_\perp$. This directly arises from the setting $R_\text{sym}(0) = z(0)$ and recognizing that $\hat{h}_{\perp+}(0) = \hat{h}_{+\perp}(0)$ by definition.

The expression for $R_\text{sym}(t)$ Eq.~\eqref{eq:R_sym} is a linear combination of four basis real timeseries, which in general are independent from each other. Note that this is not the case for the ``standard'' test Eq.~\eqref{eq:R_std}, where only two independent timeseries are needed to describe the SNR timseries.
This is the direct consequence of the fact that in the ``standard'' case the signal model only depends on a single timeseries $\hat{h}^R_{22}$, while in the general case, the two timeseries $h_+, h_\times$ are needed to specify the signal model.
We can gain more insight by redefining a different set of basis for $R_\text{sym}(t)$
\begin{align}
	\hat{h}^{S}_{++}(t) = \frac{1}{2} \left(\hat{h}_{++}(t) + \hat{h}_{\perp\perp}(t)\right) \\
	\hat{h}^{A}_{++}(t) = \frac{1}{2} \left(\hat{h}_{++}(t) - \hat{h}_{\perp\perp}(t)\right) \\
	\hat{h}^{S}_{+\perp}(t) = \frac{1}{2} \left(\hat{h}_{+\perp}(t) - \hat{h}_{\perp+}(t)\right) \\
	\hat{h}^{A}_{+\perp}(t) = \frac{1}{2} \left(\hat{h}_{+\perp}(t) + \hat{h}_{\perp+}(t)\right) 
\end{align}
and the predicted SNR timeseries takes a strikingly simple expression:
\begin{align}
	R_\text{sym}(t) =& z(0)  \left(\hat{h}^{S}_{++}(t) + i \hat{h}^{S}_{+\perp}(t) \right) \nonumber \\
		& + z^*(0)  \left(\hat{h}^{A}_{++}(t) + i \hat{h}^{A}_{+\perp}(t) \right) \label{eq:R_sym_amazing}
\end{align}
We can write Eq.~\eqref{eq:R_sym_amazing} more compactly by introducing the two complex timeseries
\begin{equation}\label{eq:complex_h_S_h_A}
	\hat{h}^{S/A}(t) =\hat{h}^{S/A}_{++}(t) + i \hat{h}^{S/A}_{+\perp}(t)
\end{equation}
so that 
\begin{equation}\label{eq:R_sym_compact}
	R_\text{sym}(t) = z(0)\,\hat{h}^{S}(t) + z^*(0)\,\hat{h}^{A}(t)
\end{equation}
The expected SNR timeseries $R_\text{sym}(t)$ can be compared to complex ``symphony'' SNR timeseries to  yield a novel generally applicable signal-consistency test $\xi_\text{sym}^2$.
We can compute the normalization of the $\xi^2$ statistics with a simple computation of the expected value of the residual timeseries $\left|z(t) - R_\text{sym}(t)\right|^2$ over different noise realizations:
\begin{align}
	<\xi^2_\text{sym}(t)> = 2 - 2\left|\hat{h}^S(t)\right|^2 + 2\left|\hat{h}^A(t)\right|^2  \label{eq:expected_value_sym}
\end{align}
The expression generalizes Eq.~\eqref{eq:expected_value_std} to the ``symphony'' search statistics.
The details of the computation are reported in App.~\ref{app:expectation}.

The first term in Eq.~\eqref{eq:R_sym_amazing} has the same structure of the ``standard'' test Eq.~\eqref{eq:R_std}. However, an additional term proportional to $z^*(0)$ enters the expression, thus breaking the convenient factorization between a complex trigger and a complex template-dependend timeseries. 
Of course, the expression agrees with the ``standard'' test for aligned-spin limit where $\tilde{\hat{h}}_+ = i\tilde{\hat{h}}_\times$. In that case, it is easy to show that $\hat{h}_\perp  = \hat{h}_\times = \hat{h}^I_{22}$ and that 
\begin{align}
	&\rescalar{\hat{h}_+}{\hat{h}_+}(t) = \rescalar{\hat{h}_\times}{\hat{h}_\times}(t) \label{eq:symmetry_22_round1}\\
	&\rescalar{\hat{h}_+}{\hat{h}_\times}(t) = - \rescalar{\hat{h}_\times}{\hat{h}_+}(t) \label{eq:symmetry_22_round2}
\end{align}
For this reason, both $\hat{h}^{A}_{++}(t)$ and  $\hat{h}^{A}_{+\perp}(t)$ are identically vanishing timeseries and Eq.~\eqref{eq:R_sym_amazing} reduces to the ``standard'' case.

\subsection{Approximating the new $\xi^2$ test} \label{sec:chisq_generalized}

Although the timeseries $R_\text{sym}(t)$ offers the best prediction for the ``symphony" SNR, it cannot be expressed as a product of the trigger $z(0)$ and a template-dependent complex timeseries. 
As discussed earlier, such a convenient factorization is crucial for reducing the computational cost of the consistency test and for deploying the test with minimal changes to existing infrastructures. Therefore, we seek an expression $R_\text{mix}(t)$ for the predicted symphony SNR timeseries that retains this convenient factorization while providing satisfactory accuracy. While this expression is only an approximation to $R_\text{sym}(t)$, it may prove adequate in certain cases.

In particular, in the case of a precessig and HM binary system, the symmetries in Eqs.~(\ref{eq:symmetry_22_round1}-\ref{eq:symmetry_22_round2}) are violated by only a ``small amount''. More formally, we observe that the magnitude of $\hat{h}^{A}_{++}(t)$ and  $\hat{h}^{A}_{+\perp}(t)$ is small in most of the practical cases and it makes sense to discard from Eq.~\eqref{eq:R_sym_amazing} the term $\propto z^*(0)$.
This leads to the an approximation $R_\text{mix}(t)$ of the predicted symphony timeseries:
\begin{align}\label{eq:R_mix}
	R_\text{mix}(t) = \, z(0) \, \left(\hat{h}^{S}_{++}(t) + i \hat{h}^{S}_{+\perp}(t) \right)
\end{align}
Therefore, we can introduce the additional ``mixed" signal-consistency test $\xi_\text{mix}^2$, which is obtained by comparing $R_\text{mix}$ with the ``symphony" SNR Eq.~\eqref{eq:sym_snr}.
The normalization factor $<\xi^2(t)>$ can be straightforwardly computed by setting $\hat{h}^A(t) = 0$ in Eq.~\eqref{eq:expected_value_sym}:
\begin{equation}
	<\xi^2_\text{sym}(t)> = 2 - 2\left|\hat{h}^S(t)\right|^2
\end{equation}
The test is equivalent to the ``standard'' $\xi^2$ Eq.~\eqref{eq:R_std_bis} with the minimal replacement $\scalar{\hat{h}^\text{R}_{22}}{\hat{h}^\text{R}_{22}}(t) \to \hat{h}^{S}(t)$.

To obtain this expression we discarded from the predicted timeseries the complex quantity $z^*(0) \hat{h}^{A}(t)$. For this reason, the order of magnitude of $\hat{h}^{A}(t)$ is intimately connected with the error introduced by the ``mixed'' predicted timeseries, hence with the performance of $\xi_\text{mix}^2$. In what follows we will use the magnitude of the peak $\rho^A$ of the $\hat{h}^{A}(t)$ as a primary indicator of the goodness of the ``mixed'' consistency test
\begin{equation}
	\rho^A = \max_t \left| \hat{h}^{A}(t) \right|
\end{equation}
As already noted, $\hat{h}^{A}(t)$ is identically zero for ``standard'' systems, hence $\rho^A$ can be also used as a metric to quantify the amount of precession and/or HM of a template.
Note that by definition $\max_t \left| \hat{h}^{S}(t) \right| = 1$, hence the quantity above authomatically measures the ratio between $\hat{h}^{A}$ and $\hat{h}^{S}$.

%

The choice of discarding the term $z^*(0) \hat{h}^{A}(t)$ is very natural but somehow arbitrary. Other choices for the residuals are possible, thus leading to a different expression for the predicted timeseries, hence different values of $\xi^2$.
A convenient alternative choice for an approximate signals consistency test consists in neglecting from $R_\text{mix}(t)$ all the terms $\mathcal{O}(\hat{h}_{+\times})$:
\begin{align}
	R_\text{mix-bis}(t) \simeq & \, z(0) \, \left[\frac{1}{2} \left(\hat{h}_{++}(t)+\hat{h}_{\times\times}(t)\right) \right. \nonumber \\
		&\left. + i \frac{1}{2} \left( \hat{h}_{+\times}(t) - \hat{h}_{\times+}(t) \right) \right] \label{eq:R_mix_gstlal}
\end{align}
The expression has the merit of being more physically interpretable than Eq.~\eqref{eq:R_mix}, as it only depends on the physical polarizations, and it defines an additional test, labeled $\xi^2_\text{mix-bis}$.
As we will see in the next section, the values of $\xi^2$ obtained with the latter expression do not significantly differ from the values of $\xi_\text{mix}^2$ obtained with Eq.~\eqref{eq:R_mix}, hence an experimenter interested in interpretability could freely use $\xi^2_\text{mix-bis}$ instead of $\xi_\text{mix}^2$.

We close by noting that by assuming the {\it approximate} symmetries Eq.~(\ref{eq:symmetry_22_round1}-\ref{eq:symmetry_22_round2}), $R_\text{mix-bis}(t)$ has the simple expression
\begin{equation}\label{eq:R_mix_gstlal_simple}
	R_\text{mix-bis}(t) \simeq z(0) \, \left(\hat{h}_{++}(t) + i \hat{h}_{+\times}(t) \right)
\end{equation}
where we can straighforward recognize the ``standard'' test with the natural replacement ${\hat{h}^\text{R}_{22} \to \hat{h}_+}$ and ${\hat{h}^\text{I}_{22} \to \hat{h}_\times}$. This is the expression that one could have naively guessed without thorough computations.

In Fig.~\ref{fig:snr_timeseries}, we present an example demonstrating how accurately the predicted SNR timeseries aligns with the actual one, comparing the ``standard," ``mixed," and ``symphony" cases. We compute the SNR timeseries obtained for precessing BBH signal with zero noise and filter the data with the same signal; we plot the measured SNR timeseries $\rho_\text{std}(t)$ and $\rho_\text{sym}(t)$ and the expected absolute values of the timeseries $R_\text{std}(t)$, $R_\text{mix}(t)$ and $R_\text{sym}(t)$.
We note that $R_\text{std}(t)$ and $\rho_\text{std}(t)$ show poor agreement with each other, while $R_\text{sym}(t)$ perfectly models the complicated features of $\rho_\text{sym}(t)$. Moreover, $R_\text{mix}(t)$ provides a satisfactory approximation to $\rho_\text{sym}(t)$, accurate to a few percents.

\begin{figure*}[t!]
	\includegraphics{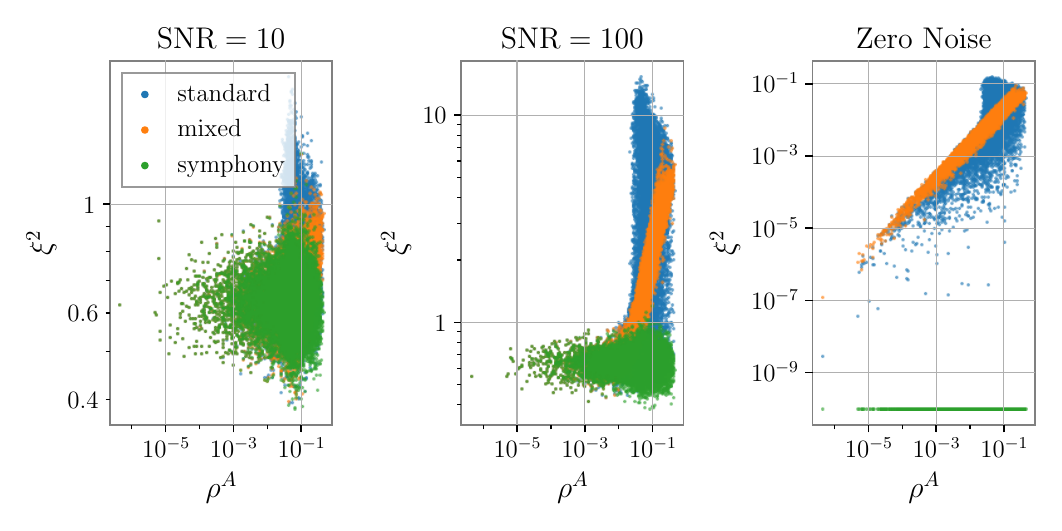}
	\caption{
	Values of $\xi^2$ Eq.~\eqref{eq:chisq} as a function of the absolute value of $(\hat{h}_+|\hat{h}_\times)$, which quantifies the precession and/or HM content of a signal. Each value is computed on random {\it precessing} BBHs, injected into Gaussian noise at a constant SNR. The left and center panels show SNRs of $20$ and $100$ respectively, while the right panel tackles the case of zero noise with injected signal normalized to 1.
	$\xi^2$ is computed using three different prescriptions.
	$\xi_\text{std}^2$ is obtained from $R_\text{std}(t)$ and $z_\text{std}(t)$ (label ``standard''). $\xi_\text{sym}^2$ is computed using $R_\text{sym}(t)$ and $z_\text{sym}(t)$ (label ``symphony"), while $\xi_\text{mix}^2$ uses $R_\text{mix}(t)$ and $z_\text{sym}(t)$.
	For this study we set ACL $ = 701$.
	}
	\label{fig:chisq_comparison}
\end{figure*}

\section{Validity and limitations of different signal consistency tests} \label{sec:validation}

In this section we study in depth the validity and the range of applicability of the various signals consistency tests discussed above, namely the new $\xi_\text{sym}^2$, the ``standard'' $\xi_\text{std}^2$ and the approximated test $\xi_\text{mix}^2$ and its alternative expression $\xi^2_\text{mix-bis}$.
After presenting a study of the capabilities of the different tests, we study the performance of the ``standard'' and ``mixed'' tests as a function of the region of the parameter space.
We also study the differences between the two alternative approximations of the ``symphony'' test $\xi_\text{mix}^2$ and $\xi_\text{mix-bis}^2$ and how the test depend on the choice of the integration window ACL.
Finally, we compare the results obtained by performing the test in Gaussian noise with those obtained with real interferometer data: this study is crucial to test the robustness of the test in a ``real life'' scenario.

To carry out our analysis, we compute $\xi_\text{std}^2$, $\xi_\text{mix}^2$, $\xi_\text{mix-bis}^2$ and $\xi_\text{sym}^2$ for $15000$ randomly sampled BBH signals, injected into Gaussian noise at different values of SNR.
We uniformly sample the total mass ${M\in[10,50]\SI{}{M_\odot}}$ and mass ratio ${q = {m_1}/{m_2} \in[1,15]}$, while reference phase and inclination, as well as the sky location, are drawn from a uniform distribution on the sphere.
We also sample the starting frequency $f_\text{min}$ in the range $[5,20]\SI{}{Hz}$.
Conveniently, we can introduce the spin tilt angle $\theta$, defined as
\begin{equation}
	\theta_i = \arccos\frac{s_\text{iz}}{s_i}
\end{equation}
where $s_i$ is the magnitude of the i-th spin. The tilt angle measures the mis-alignment of the each spin with the orbital momentum and thus it is crucial to control the amount of precession in a system, with a maximally precessing system having $\theta \simeq \pi/2$.

In our study, we explore two scenarios. In one case, we focus on precessing systems with both spins sampled isotropically inside the unit sphere. In the other case, we consider aligned-spin systems but include higher modes (HMs) in the waveform. For the latter experiment, we sample both z-components of the spins uniformly between $[-0.99,0.99]$ and we consider the HM with $(\ell, |m|) = (2,2), (2, 1), (3, 3), (3, 2), (4, 4)$.
We utilize the frequency domain approximants \texttt{IMRPhenomXP} \cite{Pratten:2020ceb} and \texttt{IMRPhenomXHM} \cite{Garcia-Quiros:2020qpx} for the two scenarios respectively.
We employ the PSD computed over the first three month of the third observing run at the LIGO Livingston detector \cite{O3_PSDs} and sample $\SI{100}{s}$ of Gaussian noise at a sample rate of $\SI{4096}{Hz}$ for each signal under study. To study the performance in real noise, we repeat the experiment by using segments of real publicly available real data~\cite{KAGRA:2023pio}, as we discussed with more details below.

\begin{figure*}[t!]
	\includegraphics{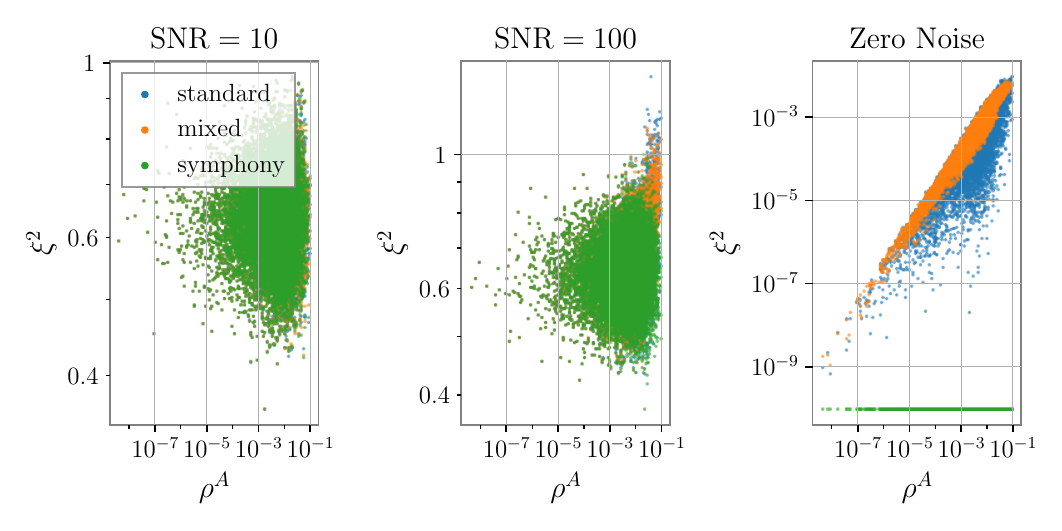}
	\caption{
	Values of $\xi^2$ Eq.~\eqref{eq:chisq} as a function of the absolute value of $(\hat{h}_+|\hat{h}_\times)$, which quantifies the precession and/or HM content of a signal. Each value is computed on random {\it aligned-spin} BBHs with HM content, injected into Gaussian noise at a constant SNR. The left and center panels show SNRs of $20$ and $100$ respectively, while the right panel tackles the case of zero noise with injected signal normalized to 1.
	$\xi^2$ is computed using three different prescriptions.
	$\xi_\text{std}^2$ is obtained from $R_\text{std}(t)$ and $z_\text{std}(t)$ (label ``standard''). $\xi_\text{sym}^2$ is computed using $R_\text{sym}(t)$ and $z_\text{sym}(t)$ (label ``symphony"), while $\xi_\text{mix}^2$ uses $R_\text{mix}(t)$ and $z_\text{sym}(t)$.
	For this study we set ACL $ = 701$.
	}
	\label{fig:chisq_comparison_HM}
\end{figure*}

\subsection{How does the different tests compare to each other?}

We compare here the performance of the different tests for different SNR summarizing the main results of the analysis described above.
In figure Fig.~\ref{fig:chisq_comparison}, we present results pertaining precessing systems, while results in Fig.~\ref{fig:chisq_comparison_HM} refer to aligned-spin system with HM.
In both figures for varying SNR, we plot the $\xi^2$ values as a function of the peak $\rho^A$ of $\hat{h}^{A}(t)$, which as discussed above is an excellent measure of the content of precession and/or HM content in a template.
Note that the $\xi^2$ values do not depend exclusively on $\rho^A$ but also on the details of the residuals $\hat{h}^{A}(t)$. Nevertheless, $\rho^A$ still remains a useful scalar quantity to quantify the lack of orthogonality of the two templates.
In the ``zero noise'' case the injected signals are normalized to one: this arbitrary choice only affects the value of $\xi^2$ with an overall scaling but it does not alter the distribution of values.
We compute the $\xi^2$ using a window (ACL) of $701$ points centered around the injection time.

As long as zero noise is considered, we note that $\xi_\text{std}^2$ and $\xi_\text{mix}^2$ are both non-zero. This means that $R_\text{std}(t)$ and $R_\text{mix}(t)$ are not able is not able to predict exactly the behaviour of the SNR timeseries.
This is expected, since in the precessing/HM regime, they are both approximations to the true SNR.
On the other hand, $\xi_\text{sym}^2$ is always zero (up to numerical noise), showing that the newly introduced $\xi_\text{sym}^2$ is the {\it optimal} test to search for generic BBH signals: this is also manifest in Fig.~\ref{fig:snr_timeseries}.

\begin{figure*}[t!]
	\includegraphics{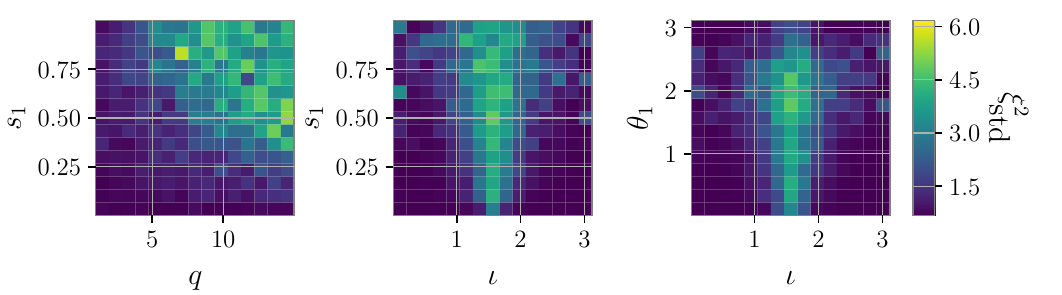}
	\caption{Performance of the ``standard'' test as a function of the parameter space. We color each bin according to the median value of $\xi_\text{std}^2$ and we consider the mass ratio $q$, the polar spin components $s_1$, $\theta_1$ and of the inclination angle $\iota$ of the 15000 test BBHs described in the text. 
	}
	\label{fig:std_performance}
\end{figure*}

\paragraph{Precessing templates}
By looking at the injected {\it precessing} signals in Fig.~\ref{fig:chisq_comparison} in the presence of noise, we see that $\xi_\text{sym}^2$ is always superior to the ``standard'' test, with an improvement as large as two orders of magnitude in the SNR $= 100$ case.
As long as $\xi_\text{mix}^2$ is considered, we observe that $\xi_\text{mix}^2$ and $\xi_\text{sym}^2$ show similar dispersions in the low SNR case. Therefore, in the presence of a substantial amount of noise, the accuracy improvement provided by $\xi_\text{sym}^2$ is negligible over the approximation given by $\xi_\text{mix}^2$.
The discrepancy between the two $\xi^2$ tests increases for $SNR = 100$; in that case, the noise level is lower and it must be of the same order of magnitude of the terms neglected to obtain $\xi_\text{mix}^2$.
We also note that the values $\xi_\text{mix}^2$ are very well correlated with $\rho^A$, thus confirming the usefulness of the latter to predict the failure of the ``mixed'' signal consistency test. 

The fact that $\xi_\text{mix}^2$ degrades its performance at high SNR should not be of concern, as the signal-consistency test is less crucial for the high SNR region. Indeed, due to the rarity of very loud signals, it is feasible to perform targeted follow up and {\it ad hoc} studies, hence assessing the significance of a trigger with other strategies.
For this reason, we conclude that $\xi_\text{mix}^2$ is likely to perform close to optimality for the vast majority of the practical applications and we recommend its implementation in any pipeline aiming to search for precessing signals.

In closing, we note that in the $\xi_\text{std}^2$ computation, we could have used Eq.~\eqref{eq:std_snr} instead of Eq.~\eqref{eq:std_snr_bis} to filter the data and, similarly, Eq.~\eqref{eq:R_std} rather than Eq.~\eqref{eq:R_std_bis} to compute the autocorrelation.
Even though in the ``standard'' case, the two expressions agree, they do not agree when precessing and/or HM template are considered.
In that case, a straightforward generalization to precession $\hat{h}^\text{R}_{22} \to \hat{h}_+$ and $\hat{h}^\text{I}_{22} \to \hat{h}_\times$ would lead $\xi_\text{std}^2$ to also take into account both polarization, with potential improvements on the efficacy of the test.
However, as the GstLAL pipeline implements the test using Eq.~\eqref{eq:R_std_bis}, we made the choice to describe the current situation.


\paragraph{HM templates}
The picture outline above changes when aligned-spin HM templates are considered in Fig.~\ref{fig:chisq_comparison_HM}. In this case, the performance of $\xi_\text{std}^2$, $\xi_\text{mix}^2$ and $\xi_\text{sym}^2$ are very comparable in the low SNR case. In the high SNR case, $\xi_\text{sym}^2$ retains a slightly better performance but the use of $\xi_\text{mix}^2$ does not bring any additional improvement over the ``standard'' test $\xi_\text{std}^2$.
Therefore, if only HMs are considered, our results indicate that $\xi_\text{std}^2$ already delivers close to optimal results, suggesting that no updates of the ``standard'' signal consistency test are required to tackle only the aligned-spin HM case.

\subsection{When does the ``standard'' test fail?}

To study the limitation of the ``standard'' $\xi^2$, in Fig.~\ref{fig:std_performance} we report the values of $\xi_\text{std}^2$ as a function of the template parameters. We only focus on the {\it precessing case}, as for the HM case $\xi_\text{std}^2$ Fig.~\ref{fig:chisq_comparison_HM} shows a good performance across the whole parameter space.
Our results suggest that the performance of the test decreases for large values of the mass ratio, large values of spin and for large spin misalignment. Moreover, precession is more visible for systems observed with a close to edge-on inclination, i.e. $\iota \simeq \pi/2$.
These findings align with the literature on the detectability of precession in BBH~\cite{CalderonBustillo:2016rlt, Fairhurst:2019srr, Green:2020ptm, McIsaac:2023ijd}, which demonstrates that precession is more detectable for asymmetric, heavily spinning edge-on systems.
The GW signals emitted by such heavily precessing acquires a more complicated structure, which translates into the lack of symmetry between the two polarizations, which fail to satisfy Eq.~\eqref{eq:symmetry_polarization_plus_cros}. As the ``standard'' test relies on such symmetry and uses only the plus polarization to predict the SNR timeseries, a large violation of Eq.~\eqref{eq:symmetry_polarization_plus_cros} also leads straighforwardly to large values of $\xi_\text{std}^2$.

\subsection{How does the ``mixed'' test perform?}

\begin{figure*}[t!]
	\includegraphics{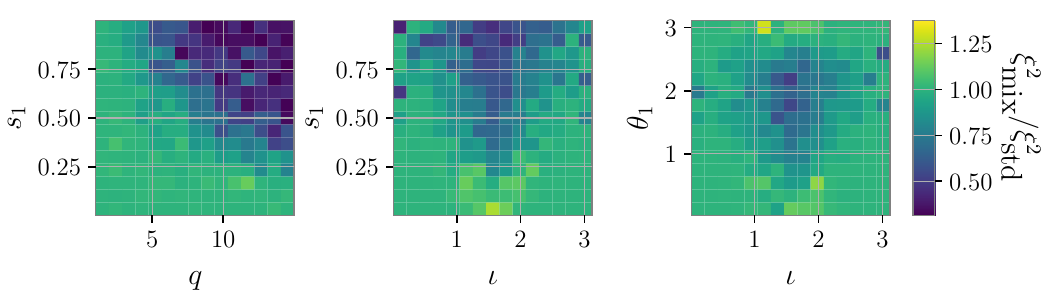}
	\caption{Comparison between the ``mixed'' and the ``standard'' signal consistency tests as a function of the parameter space. We color each bin according to the median value of $\xi_\text{mix}^2/\xi_\text{std}^2$ and we consider the mass ratio $q$, the polar spin components $s_1$, $\theta_1$ and of the inclination angle $\iota$ of the 15000 test BBHs described in the text.}
	\label{fig:mixed_improvement_over_std}
\end{figure*}

\begin{figure*}[t!]
	\includegraphics{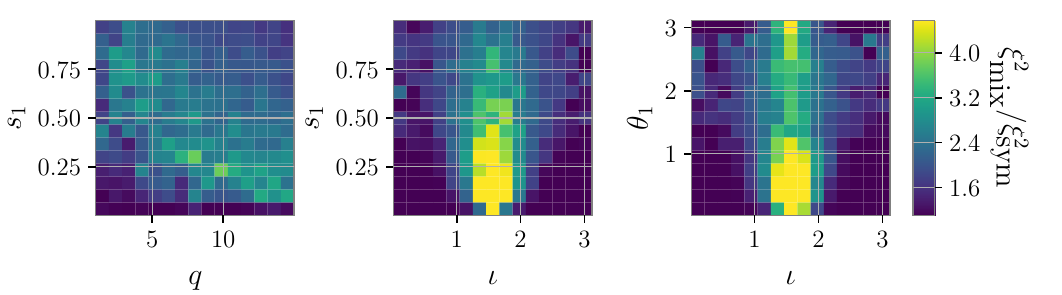}
	\caption{Comparison between the ``mixed'' and the ``standard'' signal consistency tests as a function of the parameter space. We color each bin according to the median value of $\xi_\text{mix}^2/\xi_\text{sym}^2$ and we consider the mass ratio $q$, the polar spin components $s_1$, $\theta_1$ and of the inclination angle $\iota$ of the 15000 test BBHs described in the text.}
	\label{fig:mixed_performance}
\end{figure*}

\begin{figure*}[t!]
	\includegraphics{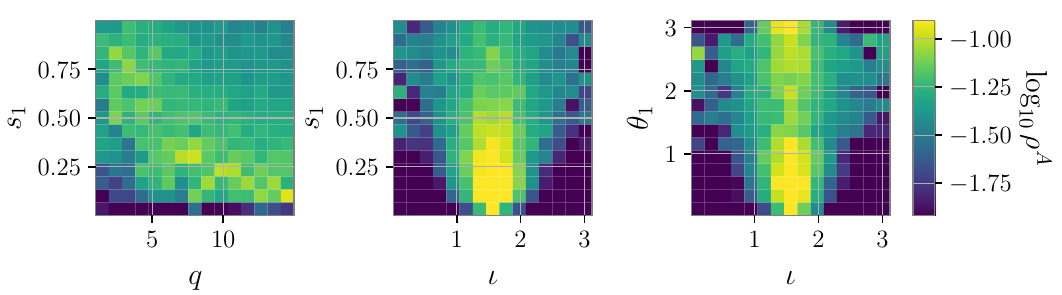}
	\caption{Values of $\rho^A$ as a function of the parameter space. We color each bin according to the median value of $\rho^A$ and we consider the mass ratio $q$, the polar spin components $s_1$, $\theta_1$ and of the inclination angle $\iota$ of the 15000 test BBHs described in the text.
	$\rho^A$ is intended as a measure of the goodness of the ``mixed'' signal consistency test and as such, it correlate very well with the values of $\xi_\text{mix}^2$ (see also Fig.~\ref{fig:mixed_performance}).}
	\label{fig:rho_A_scatter}
\end{figure*}

As we turn our attention to the performance of the approximate ``mixed'' signal consistency test, we are interested in (i) identifying the regions of the parameter space where this test provides an advantage over the ``standard'' test and (ii) evaluating the extent of performance loss compared to the optimal ``symphony'' test.
We limit our study to precessing signals, since for aligned-spin HM systems the three tests show very similar performance.

To answer the first question, we report in Fig.~\ref{fig:mixed_improvement_over_std} the ratio $\xi_\text{mix}^2/\xi_\text{std}^2$ between the ``mixed'' and the ``standard'' test, evaluated at SNR $=100$ as a function of the template paramaters.
Unsurprisingly the ``mixed'' test outperforms the ``standard'' test for system that show a strong amount of precession: in these regions of the parameter space the ``standard'' test shows poor performance while the ``mixed'' test is able to better predict the behaviour of the SNR timeseries.

\begin{figure}[t!]
	\includegraphics{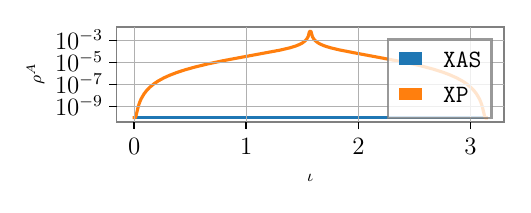}
	\caption{Numerical stability of the approximant \texttt{IMRPhenomXP} in the aligned-spin limit as a function of the inclination angle $\iota$.
		We compute the values of $\rho^A$ and for a BBH with total mass $M =  10 \SI{}{M_\odot}$ and mass ratio $q = 8$ and with spins $s_\text{1z}, s_\text{2z} = -0.4, 0.6$, seen at different inclination angles $\iota$.
		We repeat the experiment with both the precessing approximant \texttt{IMRPhenomXP} and the aligned-spin approximant \texttt{IMRPhenomXAS}. While for the aligned-spin approximant $\rho^A$ are both zero up to numerical precision, for the precessing approximant the two quantities become non zero as the inclination gets close to $\iota \simeq \pi/2$. In this case, since ${h}_\times \to 0$, the numerical noise introduced by the spin twist procedure affects the normalized cross polarization $\hat{h}_\times$.
	}
	\label{fig:rho_A_vs_iota}
\end{figure}

To understand the performance loss of the ``mixed'' test as compared to the optimal ``symphony'' test, we turn our attention to Fig.~\ref{fig:mixed_performance}, where we report the ratio $\xi_\text{mix}^2/\xi_\text{sym}^2$, also evaluated at SNR $=100$, as a function of the template paramaters.
By comparing the performance of the $\xi_\text{mix}^2$ with $\xi_\text{sym}^2$ it striking to note that the ``mixed'' test has poor performance for systems with $\iota \simeq \frac{\pi}{2}$. This is somehow expected as it is well known that the inclination increases the precession induced amplitude and phase modulation on the waveform. However, it is surprising that the worst performance occurs mostly for systems with low precession content, i.e. with $s_1 \lesssim 0.25$ and for $\theta_1 \lesssim 1$. This is puzzling because the precession content of such systems is expected to be small, due to low spins and mild mis-alignment.
We make the hyphothesis that the effect can be explained by studying the details of the {\it spin twist} procedure~\cite{Schmidt:2010it, Schmidt:2012rh, Hannam:2013oca} employed to model the precessing effects in the approximant in use.
According to this procedure, the waveform is generated with a time dependent rotation of the waveform emitted by a corresponding aligned-spin system. 
The spin twist might introduce some numerical noise in the polarizations, even in the aligned-spin limit. For $\iota$ far from $\frac{\pi}{2}$, the numerical noise is expected to be much smaller than the waveform itself, however as $h_\times \to 0$ for $\iota \to \frac{\pi}{2}$ the noise might become dominant and affect the computation of $\hat{h}^S(t)$.
A numerical study reported in Fig.~\ref{fig:rho_A_vs_iota} shows that even in the {\it aligned-spin} case the precessing approximant \texttt{IMRPhenomXP} returns a non-zero value for $\rho^A$, while the aligned-spin approximant \texttt{IMRPhenomXAS}~\cite{Pratten:2020fqn} returns the zero value consistent with the theory. For this reason, as \texttt{IMRPhenomXP} is built upon \texttt{IMRPhenomXP}, we may conclude that the spin twist procedure introduces spurious noise into the polarizations. The numerical noise becomes visible in the timeseries $\hat{h}^A(t)$ only when the cross polarization tends to zero, which happens for close to aligned-spin systems but not for heavily precessing systems: that explains why the issue is seen only for small values of $s_1$.
Further investigations should corroborate this hypothesis.

These limitations for $\iota \simeq \frac{\pi}{2}$, most likely due to the approximant employed, should not be of concern. First of all, they affect a region of the parameter space well covered by current searches and for which the ``standard'' test is mostly suitable. Secondly, future precessing approximants might solve the pathological behaviour observed in Fig.~\ref{fig:rho_A_vs_iota}, thus potentially improving the performance of the ``mixed'' test.
Finally, we note that $\xi_\text{mix}^2$ differs from $\xi_\text{sym}^2$ by at most a factor of $\sim 2$. It is left to the developer of a search to quantify the consequent loss in sensitivity and to consider whether this is an acceptable loss. Here we limit ourself to stress that $\xi_\text{mix}^2$ is more suitable to deal with the heavily precessing regime and provides a substantial improvement over the ``standard'' test.
Moreover, we stress again that the results discussed refers to signals with a high SNR of $100$, for which a {\it ad hoc} follow up strategy can be implemeted, thus making the signal consistency test less decisive.
For signals observed with a more realistic lower SNR, the impact on the $\xi_\text{mix}^2$ values is much smaller, as shown in Fig.~\ref{fig:chisq_comparison}.

In Fig.~\ref{fig:rho_A_scatter} we report the peak value $\rho^A$ of the residual timeseries $\hat{h}^A(t)$ as a function of the template parameters. We observe that the values of $\rho^A$ are very well correlated with the values of $\xi_\text{mix}^2$ reported in Fig.~\ref{fig:mixed_performance}. This confirms the utility of $\rho^A$ as a measure of the performance of the $\xi_\text{mix}^2$.

\subsection{Which ``mixed'' test should we use?}

\begin{figure}[t!]
	\includegraphics{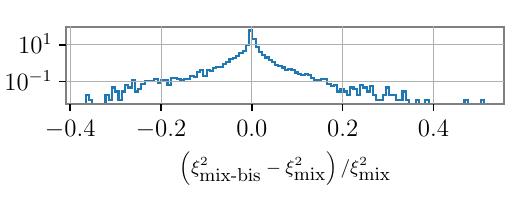}
	\caption{Discrepancies between the ``mixed'' signal consistency test $\xi_\text{mix}^2$ Eq.~\eqref{eq:R_mix} and a possible alternative definition Eq.~\eqref{eq:R_mix_gstlal} $\xi_\text{mix-bis}^2$. For the 15000 BBH signals introduced in the text, we report a histogram with the fractional difference $\frac{\xi^2_\textrm{mix-bis} - \xi^2_\textrm{mix}}{\xi^2_\textrm{mix}}$ between the two consistency tests.
	The discrepancies between the two versions are negligible, showing that Eq.~\eqref{eq:R_mix} and Eq.~\eqref{eq:R_mix_gstlal} are mostly equivalent.
	 }
	\label{fig:which_approximation}
\end{figure}

As a final analysis, we compare the performance of the two approximate tests introduced in Sec.~\ref{sec:chisq_generalized}, $\xi_\text{mix}^2$ and $\xi_\text{mix-bis}^2$. While $\xi_\text{mix}^2$ has a more straightforward definition, $\xi_\text{mix-bis}^2$ has a simpler expression and it is only written in terms of the two polarizations.
Depending on the pipeline details, an user may decide to implement either versions. For this reason, it is important to check that they give consistent results. This is done in Fig.~\ref{fig:which_approximation} where we report an histogram with the relative error of $\xi_\text{mix-bis}^2$ with respect to $\xi_\text{mix}^2$. From the histogram we learn that in $90\%$ of the cases $\xi_\text{mix-bis}^2$ has a discrepancy of less than $6\%$ from $\xi_\text{mix}^2$.
Our results show that the two expressions for the predicted SNR timeseries, Eq.~\eqref{eq:R_mix} and Eq.~\eqref{eq:R_mix_gstlal}, are mostly equivalent and they can both successfully employed in a full search, with the caveats about the validity of the approximation discussed above.

\begin{figure}[t!]
	\includegraphics{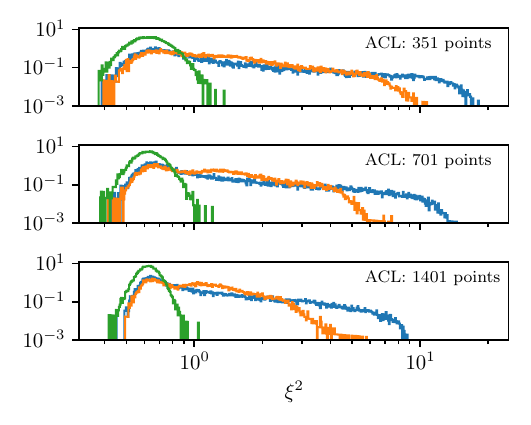}
	\caption{
	Values of $\xi^2$ for different window lengths (ACL) used for the integral in Eq.~\eqref{eq:chisq}. For each ACL, we report the values of $\xi_\text{std}^2$ (blue), $\xi_\text{mix}^2$ (orange) and $\xi_\text{sym}^2$ (green), following the color code introduced in Figs.~\ref{fig:chisq_comparison} and ~\ref{fig:chisq_comparison_HM}.
	The data refers to $15000$ {\it precessing} signals injected at $\text{SNR} = 100$.
	}
	\label{fig:ac_study}
\end{figure}

\subsection{How to choose the autocorrelation length for the test?}

In Fig.~\ref{fig:ac_study}, we study how the three $\xi^2$ tests considered depend on the choice of the integration window ACL. For brevity, we only consider the $\xi^2$ values for {\it precessing} injections with SNR $= 100$ and we choose three different values of ACL $= 351, 701, 1401$. Our results show that the $\xi^2$ values are slighly improved for a longer integration window, i.e. for a larger ACL.
However, the smallness of the differences suggests that the choice of ACL is not crucial and that overall the test is robust against different ACL values.
We then recommend to choose ACL $= 701$, which seems a satisfying trade-off between test performance and computational cost.

\subsection{How does real noise affect the test performance?}

\begin{figure}[t!]
	\includegraphics{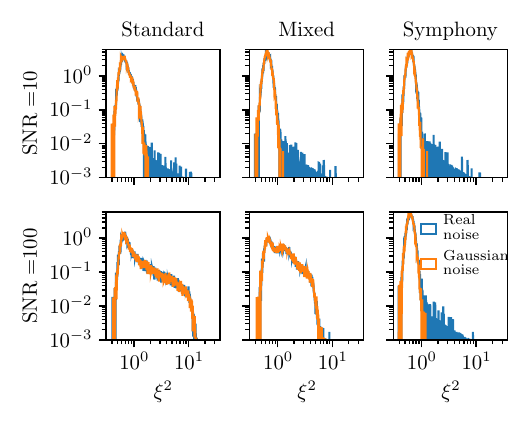}
	\caption{Performance in real detector's noise of the three signal-consistency tests discussed in the paper. We randomly select a number of precessing templates as in Fig.~\ref{fig:chisq_comparison} and we inject a corresponding signal into both Gaussian noise and real detector's noise. The detector's noise was recorded by the LIGO-Hanford observatory between GPS times $1245708288 \SI{}{s}$ and $1246756864 \SI{}{s}$.
	In each histogram, we report the values of $\xi^2$ for the Gaussian noise (orange) and Real noise (blue) case. Each panel refers to a different type of signal-consistency test and to different values of the injected SNR.
	}
	\label{fig:comparison_real_noise}
\end{figure}

We conclude our analysis by studying the performance of the various signal consistency tests in {\it real detector's noise}.
This study allows us to study a realistic scenario, where non-Gaussian artifacts may negatively impact our ability to predict the SNR timeseries.
To do so, we use the publicly available data~\cite{KAGRA:2023pio} taken during the third Observing Run (O3) by the LIGO-Hanford observatory between GPS times $1245708288 \SI{}{s}$ and $1246756864 \SI{}{s}$. We use the data to produce {\it whitened} segments of $\SI{100}{s}$, where  we inject the the test signals introduced above. We then compute the SNR timeseries, both for the ``standard'' and the ``symphony'' case, and we compute the $\xi^2$ for each injection.
We present our results in Fig.~\ref{fig:comparison_real_noise}, where for each type of signal consistency test we report the distribution of $\xi^2$ values computed in both the real and Gaussian noise cases. As above, we repeat the experiment for signals injected at an SNR of $10$ and $100$.

From our results it is manifest that the bulk of the $\xi^2$ distributions computed for both real and Gaussian noise are consistent with each other: this confirms the robustness of the test even in the real noise scenario. Morever, we note that our novel ``symphony'' produces systematically lower values of $\xi^2$  as compared to both the ``mixed'' and ``standard'' tests, further confirming its optimal performance in the high SNR case.

In some occasions the real noise produces triggers with large values of $\xi^2$ when compared with the Gaussian noise case: in these situations the non-gaussianities of the detector's output introduce artifacts in the SNR timeseries, which are not taken into account by our prediction. This produces the large values of $\xi^2$ observed in Fig.~\ref{fig:comparison_real_noise}.
This behaviour is not only expected but also desired: indeed, the $\xi^2$ test is specifically designed to discriminate between astrophysical triggers and triggers from noise artifacts. Consequently, large  $\xi^2$ values in the presence of noise artifacts indicate that the test is correctly performing its intended function.

For triggers at SNR $=100$, the ``standard'' and ``mixed'' signal consistency tests agree when computed for Gaussian and real noise: in this case, the test in Gaussian noise already produces triggers with large $\xi^2$ values, which are indistinguishable from the effect of non-Gaussian artifacts.

\section{Final remarks} \label{sec:conclusion}


We introduce a novel $\xi^2$ signal-consistency test tailored for the matched-filtering searches of gravitational waves emitted by precessing and/or higher-mode binary systems. The test measures the discrepancy between the predicted and measured SNR timeseries, as output by any matched-filtering pipeline which filters the data with a large set of templates.
While the traditional test $\xi_\text{std}^2$ is only valid for the case of aligned-spin binaries where HM are not considered, our new $\xi_\text{sym}^2$ is built upon the ``symphony" search statistics~\cite{Harry:2017weg} and it does not make any assumption about the nature of the signal to detect.
Our new $\xi_\text{sym}^2$ relies on the expression Eq.~\eqref{eq:R_sym_amazing} for the predicted matched-filtering output, which is derived here for the first time.

Thanks to the symmetry of the aligned-spin systems without HMs, the ``standard'' consistency test is convently factorized as the multiplication between a complex trigger and a complex template-dependent timeseries.
As our newly introduced test breaks such simple factorization, it requires twice the computational cost to be performed. To allieviate such cost, we also proposed an approximation $\xi_\text{mix}^2$ to our new $\xi_\text{sym}^2$, which, by respecting the simple factorization for the predicted timeseries, requires minimal changes to existing code platforms and is obtained without extra computation with respect to the ``standard'' test.

We investigated the validity of the two newly introduced tests $\xi_\text{sym}^2$ and $\xi_\text{mix}^2$, by performing an extensive study on signals injected in both Gaussian noise, reaching to four main conclusions:
\begin{itemize}
	\item The newly introduced test $\xi_\text{sym}^2$ has optimal performance for precessing and/or HM systems, being able to optimally predict the SNR timeseries in the zero noise case
	\item The traditional test $\xi_\text{std}^2$ show poor performance for heavily asymetric sytems with mis-aligned spins
	\item The approximate test $\xi_\text{mix}^2$ is very suitable for low SNR signals, while it displays some loss of performance for systems with high SNR
	\item The traditional test $\xi_\text{std}^2$ is very suitable for aligned-spin systems where HM are considered, showing similar performances to the optimal $\xi_\text{sym}^2$
\end{itemize}
These same conclusions are obtained by studing the test performance on real detector's noise, although with some outliers with high $\xi^2$ values in correspondence of loud non-gaussian transient burst of noise.

The newly introduced test, $\xi_\text{sym}^2$, can be implemented in any matched-filtering pipeline, enhancing the search for both precessing and aligned-spin signals with HM content. Additionally, the validation studies presented here will help the community better understand the limitations of the traditional $\xi_\text{std}^2$ test, benefiting any matched-filtering pipeline aimed at detecting binary black hole signals with strong precession and/or higher-mode content.

        \begin{acknowledgments}
        We thank Khun Sang Phukon for the useful remarks. We are also very grateful to the anonymous referee, who stimulated a huge improvement upon the first draft of this work.
		S.S. and S.C. are supported by the research program of the Netherlands Organization for Scientific Research (NWO). S.C. is supported by the National Science Foundation under Grant No. PHY-2309332.
		The authors are grateful for computational resources provided by the LIGO Laboratory and supported by the National Science Foundation Grants No. PHY-0757058 and No. PHY-0823459. This material is based upon work supported by NSF’s LIGO Laboratory which is a major facility fully funded by the National Science Foundation.
        \end{acknowledgments}

\newpage

\appendix
\section{Expected value of the ``symphony'' signal consistency test in Gaussian noise}\label{app:expectation}

To compute the expected value of the ``symphony'' signal consistency test in Gaussian noise we closely follow the computation presented in \cite[App.~A]{Messick:2016aqy}.
We begin by noting that since the two templates $\hat{h}_+, \hat{h}_\perp$ and the data are real timeseries, the matched filtering output $\scalar{d}{h_{+/\times}}(t)$ is also a real timeseries: ${\scalar{d}{h_{+/\perp}}(t) = \rescalar{d}{h_{+/\perp}}(t)}$.
Thus, the complex SNR timseries $z_\text{sym}(t) = \rescalar{d}{\hat{h}_+}(t) +i \rescalar{d}{\hat{h}_\times}(t)$ can be re-written as:
\begin{equation}
	z_\text{sym}(t) = \scalar{d}{\hat{h}_+ +i \hat{h}_\perp}(t)
\end{equation}

We can now compute the expect value of the residual timeseries $<\xi^2_\text{sym}(t)>$ by taking the ensemble average of $|z(t) - R_\text{sym}(t) |^2$:
\begin{align}
	<\xi^2_\text{sym}&(t)> \, = \, <\left|z(t) - z(0) \hat{h}^S(t) - z^*(0) \hat{h}^A(t) \right|^2> \nonumber \\
		&= \, - 2 \Re \Big\{ <z^*(t)z(0)> \hat{h}^S(t) \Big\}	\nonumber \\
		&- 2 \Re \Big\{ <z(t)z(0)> \hat{h}^{A*}(t) \Big\}	\nonumber \\
		&+ 2\Re \Big\{ <z(0)z(0)> \hat{h}^{*A}(t) \hat{h}^{S}(t) \Big\}	\nonumber \\
		&+ <|z(0)|^2> \left|\hat{h}^S(t)\right|^2 + <|z(0)|^2> \left|\hat{h}^A(t)\right|^2 \nonumber \\
		&+ <|z(t)|^2> \label{eq:expect_val_chisq_first_round}
\end{align}
To move forward, we need to consider some properties of Gaussian noise in frequency domain:
\begin{align}
	<\tilde{n}^*(f^\prime) \tilde{n}(f)>\,&= \frac{1}{2} S_n(f) \, \delta(f-f^\prime) \\
	<\tilde{n}(f^\prime) \tilde{n}(f)> \,&= 0
\end{align}
Using these two properties it is easy to show that
\begin{align}
	&<|z(t)|^2> \,=\, <|z(0)|^2> \,=\, 2 \\
	&<z(0)z(0)> \,=\, <z(t)z(0)>  \,=\, 0
\end{align}
by computing the relevant integrals.
Moreover, we can easily compute $<z^*(0)z(t)>$ using the definition of scalar product Eq.~\eqref{eq:time_dependent_scalar_product}:
\begin{align}
	<z^*&(0)z(t)> = 4 \int_{-\infty}^{\infty} \int_{-\infty}^{\infty} \d{f}\d{f^\prime} \; \frac{\tilde{n}^*(f^\prime)\tilde{n}(f)}{S(f^\prime)S_n(f)} \times \nonumber \\
	&\qquad\qquad\qquad \qquad
	\times (\hat{h}^*_+ -i \hat{h}^*_\perp)(\hat{h}_+ +i \hat{h}_\perp) e^{-i2\pi f t} \nonumber \\
	& = 2 \int_{-\infty}^{\infty} \d{f} \; \frac{ \hat{h}^*_+ \hat{h}_+ + \hat{h}^*_\perp \hat{h}_\perp +i (\hat{h}^*_+ \hat{h}_\perp - \hat{h}^*_\perp \hat{h}_+)}{S_n(f)} e^{-i2\pi f t} \span \nonumber \\
	& = 2 \hat{h}^{S}(t)
\end{align}

Putting everything together, we obtain a simple expression for $<\xi^2_\text{sym}(t)>$:
\begin{equation}
	<\xi^2_\text{sym}(t)> = 2 - 2  \left|\hat{h}^S(t)\right|^2 +  2 \left|\hat{h}^A(t)\right|^2
\end{equation}
which can be used to normalize the ``symphony'' $\xi^2$ test.
Note that in the ``standard'' limit, we have $\hat{h}^A(t) \to 0$ and $\hat{h}^S(t) \to \scalar{\hat{h}^\text{R}_{22}}{\hat{h}^\text{R}_{22}}(t)$, hence the two expected values $<\xi^2(t)>$ agree.

	\bibliography{biblio.bib}

\begin{thebibliography}{10}

\bibitem{LIGOScientific:2014pky}
J.~Aasi {\em et~al.}, ``{Advanced LIGO},'' {\em Class. Quant. Grav.}, vol.~32,
  p.~074001, 2015.

\bibitem{VIRGO:2014yos}
F.~Acernese {\em et~al.}, ``{Advanced Virgo: a second-generation
  interferometric gravitational wave detector},'' {\em Class. Quant. Grav.},
  vol.~32, no.~2, p.~024001, 2015.

\bibitem{KAGRA}
T.~Akutsu {\em et~al.}, ``{Overview of KAGRA: Detector design and construction
  history},'' {\em Progress of Theoretical and Experimental Physics},
  vol.~2021, p.~05A101, 08 2020.

\bibitem{LIGOScientific:2018mvr}
B.~P. Abbott {\em et~al.}, ``{GWTC-1: A Gravitational-Wave Transient Catalog of
  Compact Binary Mergers Observed by LIGO and Virgo during the First and Second
  Observing Runs},'' {\em Phys. Rev. X}, vol.~9, no.~3, p.~031040, 2019.

\bibitem{LIGOScientific:2020ibl}
R.~Abbott {\em et~al.}, ``{GWTC-2: Compact Binary Coalescences Observed by LIGO
  and Virgo During the First Half of the Third Observing Run},'' {\em Phys.
  Rev. X}, vol.~11, p.~021053, 2021.

\bibitem{LIGOScientific:2021usb}
R.~Abbott {\em et~al.}, ``{GWTC-2.1: Deep extended catalog of compact binary
  coalescences observed by LIGO and Virgo during the first half of the third
  observing run},'' {\em Phys. Rev. D}, vol.~109, no.~2, p.~022001, 2024.

\bibitem{KAGRA:2021vkt}
R.~Abbott {\em et~al.}, ``{GWTC-3: Compact Binary Coalescences Observed by LIGO
  and Virgo during the Second Part of the Third Observing Run},'' {\em Phys.
  Rev. X}, vol.~13, no.~4, p.~041039, 2023.

\bibitem{Sathyaprakash:1991mt}
B.~S. Sathyaprakash and S.~V. Dhurandhar, ``{Choice of filters for the
  detection of gravitational waves from coalescing binaries},'' {\em Phys. Rev.
  D}, vol.~44, pp.~3819--3834, 1991.

\bibitem{Dhurandhar:1992mw}
S.~V. Dhurandhar and B.~S. Sathyaprakash, ``{Choice of filters for the
  detection of gravitational waves from coalescing binaries. 2. Detection in
  colored noise},'' {\em Phys. Rev. D}, vol.~49, pp.~1707--1722, 1994.

\bibitem{Allen:2005fk}
B.~Allen, W.~G. Anderson, P.~R. Brady, D.~A. Brown, and J.~D.~E. Creighton,
  ``{FINDCHIRP: An Algorithm for detection of gravitational waves from
  inspiraling compact binaries},'' {\em Phys. Rev. D}, vol.~85, p.~122006,
  2012.

\bibitem{Cannon:2011vi}
K.~Cannon {\em et~al.}, ``{Toward Early-Warning Detection of Gravitational
  Waves from Compact Binary Coalescence},'' {\em Astrophys. J.}, vol.~748,
  p.~136, 2012.

\bibitem{Babak:2012zx}
S.~Babak {\em et~al.}, ``{Searching for gravitational waves from binary
  coalescence},'' {\em Phys. Rev. D}, vol.~87, no.~2, p.~024033, 2013.

\bibitem{Klimenko:2008fu}
S.~Klimenko, I.~Yakushin, A.~Mercer, and G.~Mitselmakher, ``{Coherent method
  for detection of gravitational wave bursts},'' {\em Class. Quant. Grav.},
  vol.~25, p.~114029, 2008.

\bibitem{Necula:2012zz}
V.~Necula, S.~Klimenko, and G.~Mitselmakher, ``{Transient analysis with fast
  Wilson-Daubechies time-frequency transform},'' {\em J. Phys. Conf. Ser.},
  vol.~363, p.~012032, 2012.

\bibitem{Drago:2020kic}
M.~Drago {\em et~al.}, ``{Coherent WaveBurst, a pipeline for unmodeled
  gravitational-wave data analysis},'' 6 2020.

\bibitem{LIGOScientific:2016sjg}
B.~P. Abbott {\em et~al.}, ``{GW151226: Observation of Gravitational Waves from
  a 22-Solar-Mass Binary Black Hole Coalescence},'' {\em Phys. Rev. Lett.},
  vol.~116, no.~24, p.~241103, 2016.

\bibitem{LIGOScientific:2017vwq}
B.~P. Abbott {\em et~al.}, ``{GW170817: Observation of Gravitational Waves from
  a Binary Neutron Star Inspiral},'' {\em Phys. Rev. Lett.}, vol.~119, no.~16,
  p.~161101, 2017.

\bibitem{Privitera:2013xza}
S.~Privitera, S.~R.~P. Mohapatra, P.~Ajith, K.~Cannon, N.~Fotopoulos, M.~A.
  Frei, C.~Hanna, A.~J. Weinstein, and J.~T. Whelan, ``{Improving the
  sensitivity of a search for coalescing binary black holes with nonprecessing
  spins in gravitational wave data},'' {\em Phys. Rev. D}, vol.~89, no.~2,
  p.~024003, 2014.

\bibitem{Adams:2015ulm}
T.~Adams, D.~Buskulic, V.~Germain, G.~M. Guidi, F.~Marion, M.~Montani,
  B.~Mours, F.~Piergiovanni, and G.~Wang, ``{Low-latency analysis pipeline for
  compact binary coalescences in the advanced gravitational wave detector
  era},'' {\em Class. Quant. Grav.}, vol.~33, no.~17, p.~175012, 2016.

\bibitem{Usman:2015kfa}
S.~A. Usman {\em et~al.}, ``{The PyCBC search for gravitational waves from
  compact binary coalescence},'' {\em Class. Quant. Grav.}, vol.~33, no.~21,
  p.~215004, 2016.

\bibitem{Capano:2016dsf}
C.~Capano, I.~Harry, S.~Privitera, and A.~Buonanno, ``{Implementing a search
  for gravitational waves from binary black holes with nonprecessing spin},''
  {\em Phys. Rev. D}, vol.~93, no.~12, p.~124007, 2016.

\bibitem{Messick:2016aqy}
C.~Messick {\em et~al.}, ``{Analysis Framework for the Prompt Discovery of
  Compact Binary Mergers in Gravitational-wave Data},'' {\em Phys. Rev. D},
  vol.~95, no.~4, p.~042001, 2017.

\bibitem{Nitz:2017svb}
A.~H. Nitz, T.~Dent, T.~Dal~Canton, S.~Fairhurst, and D.~A. Brown, ``{Detecting
  binary compact-object mergers with gravitational waves: Understanding and
  Improving the sensitivity of the PyCBC search},'' {\em Astrophys. J.},
  vol.~849, no.~2, p.~118, 2017.

\bibitem{gstlal_paper2}
S.~Sachdev, S.~Caudill, H.~Fong, R.~K.~L. Lo, C.~Messick, D.~Mukherjee,
  R.~Magee, L.~Tsukada, K.~Blackburn, P.~Brady, P.~Brockill, K.~Cannon, S.~J.
  Chamberlin, D.~Chatterjee, J.~D.~E. Creighton, P.~Godwin, A.~Gupta, C.~Hanna,
  S.~Kapadia, R.~N. Lang, T.~G.~F. Li, D.~Meacher, A.~Pace, S.~Privitera,
  L.~Sadeghian, L.~Wade, M.~Wade, A.~Weinstein, and S.~L. Xiao, ``The gstlal
  search analysis methods for compact binary mergers in advanced ligo's second
  and advanced virgo's first observing runs,'' 2019.

\bibitem{Hanna:2019ezx}
C.~Hanna {\em et~al.}, ``{Fast evaluation of multidetector consistency for
  real-time gravitational wave searches},'' {\em Phys. Rev. D}, vol.~101,
  no.~2, p.~022003, 2020.

\bibitem{Aubin:2020goo}
F.~Aubin {\em et~al.}, ``{The MBTA pipeline for detecting compact binary
  coalescences in the third LIGO\textendash{}Virgo observing run},'' {\em
  Class. Quant. Grav.}, vol.~38, no.~9, p.~095004, 2021.

\bibitem{Davies:2020tsx}
G.~S. Davies, T.~Dent, M.~T\'apai, I.~Harry, C.~McIsaac, and A.~H. Nitz,
  ``{Extending the PyCBC search for gravitational waves from compact binary
  mergers to a global network},'' {\em Phys. Rev. D}, vol.~102, no.~2,
  p.~022004, 2020.

\bibitem{Chu:2020pjv}
Q.~Chu {\em et~al.}, ``{SPIIR online coherent pipeline to search for
  gravitational waves from compact binary coalescences},'' {\em Phys. Rev. D},
  vol.~105, no.~2, p.~024023, 2022.

\bibitem{Ewing:2023qqe}
B.~Ewing {\em et~al.}, ``{Performance of the low-latency GstLAL inspiral search
  towards LIGO, Virgo, and KAGRA's fourth observing run},'' 5 2023.

\bibitem{Blackburn:2008ah}
L.~Blackburn {\em et~al.}, ``{The LSC Glitch Group: Monitoring Noise Transients
  during the fifth LIGO Science Run},'' {\em Class. Quant. Grav.}, vol.~25,
  p.~184004, 2008.

\bibitem{LIGOScientific:2016gtq}
B.~P. Abbott {\em et~al.}, ``{Characterization of transient noise in Advanced
  LIGO relevant to gravitational wave signal GW150914},'' {\em Class. Quant.
  Grav.}, vol.~33, no.~13, p.~134001, 2016.

\bibitem{Cabero:2019orq}
M.~Cabero {\em et~al.}, ``{Blip glitches in Advanced LIGO data},'' {\em Class.
  Quant. Grav.}, vol.~36, no.~15, p.~15, 2019.

\bibitem{LIGO:2020zwl}
S.~Soni {\em et~al.}, ``{Reducing scattered light in LIGO's third observing
  run},'' {\em Class. Quant. Grav.}, vol.~38, no.~2, p.~025016, 2020.

\bibitem{LIGO:2021ppb}
D.~Davis {\em et~al.}, ``{LIGO detector characterization in the second and
  third observing runs},'' {\em Class. Quant. Grav.}, vol.~38, no.~13,
  p.~135014, 2021.

\bibitem{Allen:2004gu}
B.~Allen, ``{${\chi}^{2}$ time-frequency discriminator for gravitational wave
  detection},'' {\em Phys. Rev. D}, vol.~71, p.~062001, 2005.

\bibitem{Shawhan:2004qq}
P.~Shawhan and E.~Ochsner, ``{A New waveform consistency test for gravitational
  wave inspiral searches},'' {\em Class. Quant. Grav.}, vol.~21,
  pp.~S1757--S1766, 2004.

\bibitem{Cannon:2015gha}
K.~Cannon, C.~Hanna, and J.~Peoples, ``{Likelihood-Ratio Ranking Statistic for
  Compact Binary Coalescence Candidates with Rate Estimation},'' 4 2015.

\bibitem{Nitz:2017lco}
A.~H. Nitz, ``{Distinguishing short duration noise transients in LIGO data to
  improve the PyCBC search for gravitational waves from high mass binary black
  hole mergers},'' {\em Class. Quant. Grav.}, vol.~35, no.~3, p.~035016, 2018.

\bibitem{Dhurandhar:2017aan}
S.~Dhurandhar, A.~Gupta, B.~Gadre, and S.~Bose, ``{A unified approach to
  $\chi^2$ discriminators for searches of gravitational waves from compact
  binary coalescences},'' {\em Phys. Rev. D}, vol.~96, no.~10, p.~103018, 2017.

\bibitem{Gayathri:2019omo}
V.~Gayathri, P.~Bacon, A.~Pai, E.~Chassande-Mottin, F.~Salemi, and G.~Vedovato,
  ``{Astrophysical signal consistency test adapted for gravitational-wave
  transient searches},'' {\em Phys. Rev. D}, vol.~100, no.~12, p.~124022, 2019.

\bibitem{Godwin:2020weu}
P.~Godwin {\em et~al.}, ``{Incorporation of Statistical Data Quality
  Information into the GstLAL Search Analysis},'' 10 2020.

\bibitem{McIsaac:2022odb}
C.~McIsaac and I.~Harry, ``{Using machine learning to autotune chi-squared
  tests for gravitational wave searches},'' {\em Phys. Rev. D}, vol.~105,
  no.~10, p.~104056, 2022.

\bibitem{Tsukada:2023edh}
L.~Tsukada {\em et~al.}, ``{Improved ranking statistics of the GstLAL inspiral
  search for compact binary coalescences},'' {\em Phys. Rev. D}, vol.~108,
  no.~4, p.~043004, 2023.

\bibitem{DalCanton:2014qjd}
T.~Dal~Canton, A.~P. Lundgren, and A.~B. Nielsen, ``{Impact of precession on
  aligned-spin searches for neutron-star\textendash{}black-hole binaries},''
  {\em Phys. Rev. D}, vol.~91, no.~6, p.~062010, 2015.

\bibitem{PhysRevD.89.024010}
I.~W. Harry, A.~H. Nitz, D.~A. Brown, A.~P. Lundgren, E.~Ochsner, and
  D.~Keppel, ``Investigating the effect of precession on searches for
  neutron-star--black-hole binaries with advanced ligo,'' {\em Phys. Rev. D},
  vol.~89, p.~024010, Jan 2014.

\bibitem{PhysRevD.102.041302}
S.~Fairhurst, R.~Green, M.~Hannam, and C.~Hoy, ``When will we observe binary
  black holes precessing?,'' {\em Phys. Rev. D}, vol.~102, p.~041302, Aug 2020.

\bibitem{Indik:2016qky}
N.~Indik, K.~Haris, T.~Dal~Canton, H.~Fehrmann, B.~Krishnan, A.~Lundgren, A.~B.
  Nielsen, and A.~Pai, ``{Stochastic template bank for gravitational wave
  searches for precessing neutron-star\textendash{}black-hole coalescence
  events},'' {\em Phys. Rev. D}, vol.~95, no.~6, p.~064056, 2017.

\bibitem{Harry:2016ijz}
I.~Harry, S.~Privitera, A.~Boh\'e, and A.~Buonanno, ``{Searching for
  Gravitational Waves from Compact Binaries with Precessing Spins},'' {\em
  Phys. Rev. D}, vol.~94, no.~2, p.~024012, 2016.

\bibitem{McIsaac:2023ijd}
C.~McIsaac, C.~Hoy, and I.~Harry, ``{Search technique to observe precessing
  compact binary mergers in the advanced detector era},'' {\em Phys. Rev. D},
  vol.~108, no.~12, p.~123016, 2023.

\bibitem{CalderonBustillo:2015lrt}
J.~Calder\'on~Bustillo, S.~Husa, A.~M. Sintes, and M.~P\"urrer, ``{Impact of
  gravitational radiation higher order modes on single aligned-spin
  gravitational wave searches for binary black holes},'' {\em Phys. Rev. D},
  vol.~93, no.~8, p.~084019, 2016.

\bibitem{Harry:2017weg}
I.~Harry, J.~Calder\'on~Bustillo, and A.~Nitz, ``{Searching for the full
  symphony of black hole binary mergers},'' {\em Phys. Rev. D}, vol.~97, no.~2,
  p.~023004, 2018.

\bibitem{Chandra:2022ixv}
K.~Chandra, J.~Calder\'on~Bustillo, A.~Pai, and I.~Harry, ``{First
  gravitational-wave search for intermediate-mass black hole mergers with
  higher order harmonics},'' 7 2022.

\bibitem{2021PhRvD.103b4042M}
C.~{Mills} and S.~{Fairhurst}, ``{Measuring gravitational-wave higher-order
  multipoles},'' {\em Phys. Rev. D}, vol.~103, p.~024042, Jan. 2021.

\bibitem{Wadekar:2023kym}
D.~Wadekar, T.~Venumadhav, A.~K. Mehta, J.~Roulet, S.~Olsen, J.~Mushkin,
  B.~Zackay, and M.~Zaldarriaga, ``{A new approach to template banks of
  gravitational waves with higher harmonics: reducing matched-filtering cost by
  over an order of magnitude},'' 10 2023.

\bibitem{cannon2020gstlal}
K.~Cannon, S.~Caudill, C.~Chan, B.~Cousins, J.~D.~E. Creighton, B.~Ewing,
  H.~Fong, P.~Godwin, C.~Hanna, S.~Hooper, R.~Huxford, R.~Magee, D.~Meacher,
  C.~Messick, S.~Morisaki, D.~Mukherjee, H.~Ohta, A.~Pace, S.~Privitera,
  I.~de~Ruiter, S.~Sachdev, L.~Singer, D.~Singh, R.~Tapia, L.~Tsukada,
  D.~Tsuna, T.~Tsutsui, K.~Ueno, A.~Viets, L.~Wade, and M.~Wade, ``Gstlal: A
  software framework for gravitational wave discovery,'' 2020.

\bibitem{Maggiore:2007ulw}
M.~Maggiore, {\em {Gravitational Waves. Vol. 1: Theory and Experiments}}.
\newblock Oxford Master Series in Physics, Oxford University Press, 2007.

\bibitem{Apostolatos:1994mx}
T.~A. Apostolatos, C.~Cutler, G.~J. Sussman, and K.~S. Thorne, ``{Spin induced
  orbital precession and its modulation of the gravitational wave forms from
  merging binaries},'' {\em Phys. Rev. D}, vol.~49, pp.~6274--6297, 1994.

\bibitem{Kidder:1992fr}
L.~E. Kidder, C.~M. Will, and A.~G. Wiseman, ``{Spin effects in the inspiral of
  coalescing compact binaries},'' {\em Phys. Rev. D}, vol.~47, no.~10,
  pp.~R4183--R4187, 1993.

\bibitem{Kidder:1995zr}
L.~E. Kidder, ``{Coalescing binary systems of compact objects to postNewtonian
  5/2 order. 5. Spin effects},'' {\em Phys. Rev. D}, vol.~52, pp.~821--847,
  1995.

\bibitem{Buonanno:2002fy}
A.~Buonanno, Y.-b. Chen, and M.~Vallisneri, ``{Detecting gravitational waves
  from precessing binaries of spinning compact objects: Adiabatic limit},''
  {\em Phys. Rev. D}, vol.~67, p.~104025, 2003.
\newblock [Erratum: Phys.Rev.D 74, 029904 (2006)].

\bibitem{Campanelli:2006fy}
M.~Campanelli, C.~O. Lousto, Y.~Zlochower, B.~Krishnan, and D.~Merritt, ``{Spin
  Flips and Precession in Black-Hole-Binary Mergers},'' {\em Phys. Rev. D},
  vol.~75, p.~064030, 2007.

\bibitem{Pekowsky:2012sr}
L.~Pekowsky, J.~Healy, D.~Shoemaker, and P.~Laguna, ``{Impact of higher-order
  modes on the detection of binary black hole coalescences},'' {\em Phys. Rev.
  D}, vol.~87, no.~8, p.~084008, 2013.

\bibitem{Healy:2013jza}
J.~Healy, P.~Laguna, L.~Pekowsky, and D.~Shoemaker, ``{Template Mode
  Hierarchies for Binary Black Hole Mergers},'' {\em Phys. Rev. D}, vol.~88,
  no.~2, p.~024034, 2013.

\bibitem{Finn:1992xs}
L.~S. Finn and D.~F. Chernoff, ``{Observing binary inspiral in gravitational
  radiation: One interferometer},'' {\em Phys. Rev. D}, vol.~47,
  pp.~2198--2219, 1993.

\bibitem{Jaranowski:1998qm}
P.~Jaranowski, A.~Krolak, and B.~F. Schutz, ``{Data analysis of gravitational -
  wave signals from spinning neutron stars. 1. The Signal and its detection},''
  {\em Phys. Rev. D}, vol.~58, p.~063001, 1998.

\bibitem{O3_PSDs}
{Abbott, R. and others}, ``{Noise curves used for Simulations in the update of
  the Observing Scenarios Paper}.''
  \url{https://dcc.ligo.org/LIGO-T2000012/public}, 2022.
\newblock [Online; accessed 29-January-2024].

\bibitem{Pratten:2020ceb}
G.~Pratten {\em et~al.}, ``{Computationally efficient models for the dominant
  and subdominant harmonic modes of precessing binary black holes},'' {\em
  Phys. Rev. D}, vol.~103, no.~10, p.~104056, 2021.

\bibitem{Capano:2013raa}
C.~Capano, Y.~Pan, and A.~Buonanno, ``{Impact of higher harmonics in searching
  for gravitational waves from nonspinning binary black holes},'' {\em Phys.
  Rev. D}, vol.~89, no.~10, p.~102003, 2014.

\bibitem{Schmidt:2014iyl}
P.~Schmidt, F.~Ohme, and M.~Hannam, ``{Towards models of gravitational
  waveforms from generic binaries II: Modelling precession effects with a
  single effective precession parameter},'' {\em Phys. Rev. D}, vol.~91, no.~2,
  p.~024043, 2015.

\bibitem{Garcia-Quiros:2020qpx}
C.~Garc\'\i{}a-Quir\'os, M.~Colleoni, S.~Husa, H.~Estell\'es, G.~Pratten,
  A.~Ramos-Buades, M.~Mateu-Lucena, and R.~Jaume, ``{Multimode frequency-domain
  model for the gravitational wave signal from nonprecessing black-hole
  binaries},'' {\em Phys. Rev. D}, vol.~102, no.~6, p.~064002, 2020.

\bibitem{KAGRA:2023pio}
R.~Abbott {\em et~al.}, ``{Open Data from the Third Observing Run of LIGO,
  Virgo, KAGRA, and GEO},'' {\em Astrophys. J. Suppl.}, vol.~267, no.~2, p.~29,
  2023.

\bibitem{CalderonBustillo:2016rlt}
J.~Calder\'on~Bustillo, P.~Laguna, and D.~Shoemaker, ``{Detectability of
  gravitational waves from binary black holes: Impact of precession and higher
  modes},'' {\em Phys. Rev. D}, vol.~95, no.~10, p.~104038, 2017.

\bibitem{Fairhurst:2019srr}
S.~Fairhurst, R.~Green, M.~Hannam, and C.~Hoy, ``{When will we observe binary
  black holes precessing?},'' {\em Phys. Rev. D}, vol.~102, no.~4, p.~041302,
  2020.

\bibitem{Green:2020ptm}
R.~Green, C.~Hoy, S.~Fairhurst, M.~Hannam, F.~Pannarale, and C.~Thomas,
  ``{Identifying when Precession can be Measured in Gravitational Waveforms},''
  {\em Phys. Rev. D}, vol.~103, no.~12, p.~124023, 2021.

\bibitem{Schmidt:2010it}
P.~Schmidt, M.~Hannam, S.~Husa, and P.~Ajith, ``{Tracking the precession of
  compact binaries from their gravitational-wave signal},'' {\em Phys. Rev. D},
  vol.~84, p.~024046, 2011.

\bibitem{Schmidt:2012rh}
P.~Schmidt, M.~Hannam, and S.~Husa, ``{Towards models of gravitational
  waveforms from generic binaries: A simple approximate mapping between
  precessing and non-precessing inspiral signals},'' {\em Phys. Rev. D},
  vol.~86, p.~104063, 2012.

\bibitem{Hannam:2013oca}
M.~Hannam, P.~Schmidt, A.~Boh\'e, L.~Haegel, S.~Husa, F.~Ohme, G.~Pratten, and
  M.~P\"urrer, ``{Simple Model of Complete Precessing Black-Hole-Binary
  Gravitational Waveforms},'' {\em Phys. Rev. Lett.}, vol.~113, no.~15,
  p.~151101, 2014.

\bibitem{Pratten:2020fqn}
G.~Pratten, S.~Husa, C.~Garcia-Quiros, M.~Colleoni, A.~Ramos-Buades,
  H.~Estelles, and R.~Jaume, ``{Setting the cornerstone for a family of models
  for gravitational waves from compact binaries: The dominant harmonic for
  nonprecessing quasicircular black holes},'' {\em Phys. Rev. D}, vol.~102,
  no.~6, p.~064001, 2020.

\end{thebibliography}
	\bibliographystyle{ieeetr}

\end{document}